\documentclass{osa-article}
\usepackage{colortbl}
\usepackage[bbgreekl]{mathbbol}
\usepackage{bm}
\usepackage{amssymb}
\usepackage{amsmath}
\usepackage{empheq}
\usepackage{cancel}
\usepackage{comment}

\newcommand{\widefbox}[1]{\fbox{#1}}
\newcommand{\norm}[1]{\left\lVert#1\right\rVert}

\newcommand{\citeasnoun}[1]{Ref.~\citenum{#1}}
\newcommand{\citeasnouns}[1]{Refs.~\citenum{#1}}

\newcommand{\figref}[1]{Fig.~\ref{fig:#1}}
\newcommand{\figrefbegin}[1]{Figure~\ref{fig:#1}}

\newcommand{\secref}[1]{Sec.~\ref{#1}}
\renewcommand{\eqref}[1]{Eq.~(\ref{eq:#1})}

\newcommand{\eqsref}[2]{(\ref{eq:#1},\ref{eq:#2})}

\renewcommand{\vec}[1]{\mathbf{#1}}
\newcommand{\vecg}[1]{\boldsymbol{#1}}
\newcommand{\mat}[1]{\mathbb{#1}}

\newcommand{\Scale}[2][4]{\scalebox{#1}{$#2$}}%

\journal{osac}

\articletype{Research Article}

\begin{document}

\title{Ab-initio theory of quantum fluctuations and relaxation oscillations in multimode lasers}
\author{Adi Pick,\authormark{1,*} Alexander Cerjan,\authormark{2} and Steven G. Johnson\authormark{3}}
\authormark{1}Faculty of Chemistry, Technion-Israel Institute of Technology, Haifa, Israel.\\
\authormark{2}Department of Physics, The Pennsylvania State University, University Park, Pennsylvania 16802, USA\\
\authormark{3}Department of Mathematics, Massachusetts Institute of Technology, 77 Massachusetts Avenue, Cambridge, Massachusetts 02139, USA

\email{\authormark{*}pick.adi@gmail.com} 



\begin{abstract}
We present an \emph{ab-initio} semi-analytical solution for   the  noise spectrum of complex-cavity micro-structured lasers, including central Lorentzian peaks at the multimode  lasing frequencies and additional sidepeaks due to relaxation-oscillation (RO) dynamics.  In~\citeasnoun{pick2015ab}, we computed the central-peak linewidths by solving   generalized laser rate equations, which we  derived from the Maxwell--Bloch equations by   invoking the fluctuation--dissipation theorem to   relate the noise correlations  to the steady-state lasing properties; Here, we generalize this approach and obtain  the entire laser  spectrum, focusing on the RO sidepeaks. Our formulation   treats inhomogeneity, cavity openness, nonlinearity, and multimode effects accurately. We find  a number of new effects, including new multimode RO sidepeaks and three generalized $\alpha$ factors. Last, we apply  our formulas to compute the noise spectrum of  single- and multimode  photonic-crystal lasers.
\end{abstract}



\section{Introduction \label{intro}}

The fluctuation--dissipation theorem (FDT)~\cite{Callen1951,Rytov1989,Dzyaloshinkii1961}, which relates microscopic  fluctuations to  macroscopic susceptibilities, forms  the basis of the modern understanding of electromagnetic fluctuation-based phenomena, such as  Casimir forces  and radiative heat transfer~\cite{buhmann2013dispersion,volokitin2017electromagnetic,dalvit2011casimir,reid2013fluctuation}.   In a laser, spontaneous-emission noise causes  fluctuations in the field that broaden the    emission spectrum to cover a finite bandwidth~\cite{Schawlow1958}. { A laser can be treated as a negative-temperature system at local equilibrium and a generalized  FDT can be used, in this context,   to relate the  correlations of the noise  to   the imaginary part of the dielectric permittivity~\cite{butcher1965fluctuation,Henry1996,Matloob1997,Lifshitz1980,Eckhardt1982}.} This relation produces a formula for the  noise spectrum in terms of   the laser steady-state properties~\cite{Henry1982,Henry1986,Arnaud1986,Duan1990,Exter1995}.  While traditional laser-noise  theories are excellent at predicting the  properties of macro-scale lasers~\cite{vahala1983observation,Exter1992}, they fail when applied to microstructured lasers with wavelength-scale inhomogeneities, and they also require empirical parameters~\cite{Chong2012}.   Inspired by the recent  FDT-based advances in stochastic  electromagnetism~\cite{rahi2009scattering,rodriguez2011casimir}, we recently employed  similar tools to obtain an analytic solution for the   \emph{linewidth of the central lasing peaks}~\cite{pick2015ab}, which avoids all of the traditional approximations and finds new linewidth corrections   for highly inhomogeneous and strongly nonlinear lasers.   In this paper,  we present a closed-form expression for the \emph{entire laser spectrum}, including sidepeaks that arise due to oscillations of the laser intensity  as it relaxes to the steady state following noise-driven perturbations.   Our single-mode  formula [\eqref{single-mode-result}] agrees with earlier theories in the appropriate limits (reducing to the result of \cite{Exter1992} in the limit of constant atomic-relaxation rates and to \cite{pick2015ab} when phase and intensity fluctuations of the field are decoupled) and deviates substantially  for lasers with wavelength-scale inhomogeneity.  We   predict  several  new   effects, such as enhanced smearing of the sidepeaks, new inhomogeneous corrections to the $\alpha$ factor { (which is the dominant linewidth broadening factor in semiconductor lasers~\cite{Henry1982,Osinski1987}),} and new   multimode sidepeaks  due to amplitude modulation of the relaxation-oscillation (RO)  signal.

\begin{figure}[t]
\centering
                 \includegraphics[width=0.8\textwidth]{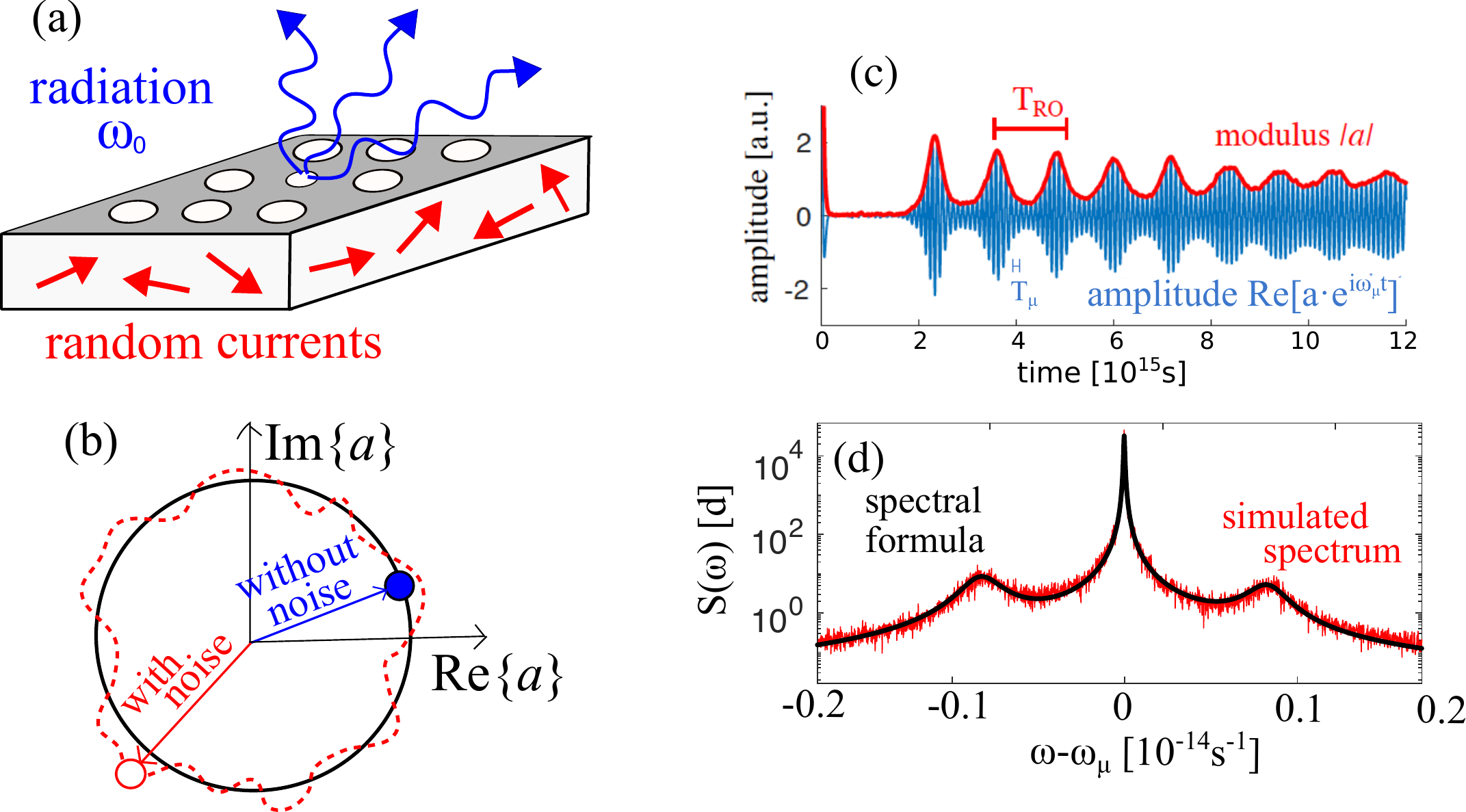}

                                    \caption{
(a) A photonic-crystal laser cavity, with stimulated emission (blue) and spontaneous emission noise  (red arrows). 
(b) Phasor diagram  of the   amplitude of a single-mode laser $a(t)$ [which obeys \eqref{Osci-single}].  Without noise, $a$ undergoes harmonic oscillations at the laser   frequency $\omega_\mu$ (black), but in the presence of noise, $a$ exhibits small intensity fluctuations and large phase drifts  (red). 
(c) Evolution of the rotated  mode  amplitude, $\mathrm{Re}[ae^{i\omega_L t}]$, (blue) and  modulus, $|a|$, (red) from the initial state until reaching the steady state. The field amplitude  oscillates at frequency $\omega_\mu$ while the  modulus undergoes relaxation oscillations with frequency  $\omega_{\mathrm{RO}}$.
(d)  Noise spectrum, obtained from the Fourier transform of the simulated solution of \eqref{Osci-single} (red) and  by evaluating the spectral formula (black) [\eqref{single-mode-result}]. }
\label{fig:RO-illustration}
  \end{figure}

Laser dynamics are surveyed in many sources~\cite{Sargent1974,Haken1984,Haken1985,Svelto1976,Milonni2010}, but it is useful to review here a simple physical picture of laser noise.  A resonant cavity [e.g., light bouncing between two mirrors or a photonic-crystal (PhC) microcavity~\cite{Jannopoulos2008} as in \figref{RO-illustration}(a)] traps light for a long time in some volume, and lasing occurs when a gain medium is ``pumped" to a population ``inversion" of excited states to the point (threshold) where gain balances loss. The nonlinear  interaction between the field  and   the gain medium   stabilizes the  system at a  steady state. If noise were absent, the field  would  perform harmonic oscillations and  the laser-power spectrum would consist of  delta functions at the oscillation frequencies, $\omega_\mu$.  However,  noise [represented by red arrows in panel (a)] is always present, and it  ``kicks'' the  field  away from the steady state.   Fluctuations in the intensity of the field are  suppressed by the nonlinear interaction with the  gain,  while   phase fluctuations  can be large [see panel (b)].  The phase undergoes  a Brownian motion, 
which  leads to broadening of the central lasing peaks~\cite{Henry1982,Lax1967a}. The effect of intensity fluctuations  depends on the relative relaxation rates of the    gain and  the field~\cite{Arecchi1984,Oppo1986,Lugiato1984}.  When the  population inversion  of the medium decays much more rapidly than the field  (a regime  called ``class-A lasers''),  intensity fluctuations decay exponentially to the steady state. A non-zero  intensity-phase coupling leads to enhanced phase variance, which increases the linewidths of the central peaks by a factor of $1+\alpha^2$~\cite{Henry1982,Henry1986,Osinski1987} (where $\alpha$ is  ``the amplitude--phase coupling'' and can be computed from the lasing mode and material properties~\cite{Henry1986,Osinski1987}).    In the  limit of comparable inversion and field  relaxation rates (i.e., in ``class-B lasers''), the inversion and  laser intensity  undergo relaxation oscillations (ROs)~\cite{Svelto1976,Milonni2010}, which produce, in addition to  central-peak broadening, a series of   sidepeaks in the noise spectrum [see panels (c--d), obtained by numerically solving   \eqref{Osci-single} and \eqref{single-mode-result}, as explained below]. The amplitudes of subsequent  peaks in the series  decrease exponentially and, in most cases, only the first-order sidepeaks are measurable.  Last, when fluctuations in the inversion relax much more slowly  than the field (i.e., in ``class-C lasers''), multimode lasing is unstable and the dynamics is chaotic~\cite{Arecchi1984}. This paper focuses on RO sidepeaks, which are relevant for  class-B lasers.

RO sidepeaks were  first predicted and measured  by  Vahala   \emph{et al.}~\cite{vahala1983observation,vahala1983semiclassicalB}. The early  measurements found an asymmetry between the amplitudes of  the blue  and red   sidepeaks~\cite{Vahala1983,Westbrook1987}.  Later work by  van Exter \emph{et al.}~\cite{Exter1992}  attributed this asymmetry to the  $\alpha$ factor.  Since most  typical  semiconductor lasers have a positive  $\alpha$ factor~\cite{Vahala1983,Westbrook1987}, this  result implied  that  the  red sidepeaks   are usually  stronger than  blue sidepeaks  (negative $\alpha$ factors are possible~\cite{Osinski1987,pereira2016linewidth}, but are less common). 
The van Exter  work used  the traditional laser   rate equations in order to derive the power-spectrum formula, but these rate equations were  derived under severe approximations and, hence, limit the generality of this  result.  In this work, we remedy this shortcoming by using generalized rate equations  [\eqref{Osci-single}], which treat the inhomogeneity and nonlinearity in the laser medium accurately. These equations  were derived in~\citeasnoun{pick2015ab}, and are introduced in the next section.

\section{From Langevin Maxwell--Bloch to the oscillator equations \label{MB2oscillator}}

The starting point of our derivation in \citeasnoun{pick2015ab} is the Langevin Maxwell--Bloch equations \cite{Haken1984,Lamb1964}, which describe the dynamics of an  electromagnetic  field ($\vec{E}$) interacting with a two-level gain medium, represented by  polarization ($\vec{P}$) and population inversion (${D}$), in the presence of noise ($\vec{F}_S$):
\begin{subequations}
\begin{gather}
\nabla\times\nabla\times \vec{E}+\varepsilon_c(\vec{x})\hspace{2pt}\ddot{\vec{E}}
=-\ddot{\vec{P}}+\vec{F}_S
\label{eq:e-dot}\\
\dot{\vec{P}}=-i(\omega_a-i\gamma_\perp)\vec{P}-i\gamma_\perp \vec{E}\, D
\label{eq:p-dot}\\
\dot{D}=-\gamma_\parallel\left[D_0\,F(\vec{x})-D+\frac{i}{2}
(\vec{E}\cdot\vec{P}^*-\vec{E}^*\cdot\vec{P})\right]
\label{eq:d-dot}
\end{gather}
\label{eq:Maxwell-Bloch}
\end{subequations}
\hspace{-2pt}
The first equation is a  Maxwell-type  equation for the field in a cavity with passive permittivity $\varepsilon_c(\vec{x})$, which is  driven by the atomic polarization  and the noise.  The second equation is an oscillator equation for the polarization, with frequency $\omega_a$ and damping rate $\gamma_\perp$, which is driven by the field and the inversion. Last,   the inversion is created    by   an external pump source [with  $D_0$  and $F(\vec{x})$ representing  the pump strength and  spatial  distribution]; it  is saturated by the field and atomic polarization,  relaxing to the steady state at a rate $\gamma_\parallel$. 
Throughout the paper, we use bold letters to denote vectors.  The units and  underlying  assumptions of this model are discussed in \cite{Tureci2006,Tureci2007,Tureci2008,Ge2010}.  
 {  
 Note that \eqref{e-dot} neglects spatial dispersion~\cite{agarwal1974electromagnetic} (i.e., nonlocal effects),  which  may arise  due to  gain diffusion~\cite{cerjan2015steady}, e.g., in  some molecular-gas~\cite{chua2011spatio} and semiconductor lasers~\cite{boardman1976surface}. Such    effects will not alter  the noise spectrum when the diffusion is  much slower than the bare inversion relaxation rate $\gamma_\parallel$; the strong-diffusion regime is beyond the scope of this work.  
For simplicity of presentation, \eqref{e-dot} neglects  also spectral dispersion (nonlocality in time) of the passive permittivity. However, our derivation of the noise spectrum is valid also for dispersive  media, so we include a frequency dependence in the Fourier transform of $\varepsilon_c(\vec{x})$, which appears in Table 1.}

{  Noise is incorporated   by  including  a fluctuating current source, $\vec{F}_S$, in the equation for the field [\eqref{e-dot}], whose correlations are given by the FDT, under the assumption of local thermal equilibrium. Although  lasers are pumped nonlinear systems, when operating at  steady state, they reach thermal equilibrium~\cite{Callen1951,Rytov1989,Dzyaloshinkii1961,Lifshitz1980,Eckhardt1982} since dissipation by optical absorption must be balanced by spontaneous emission. The probability distribution of the atomic populations obeys Boltzmann statistics, with an effective inverse temperature defined
 as~\cite{Jeffers1993,Matloob1997,Patra1999}
\begin{equation}
\beta(\vec{x})\equiv  \frac{1}{\hbar\omega_0}\ln\left(\frac{N_1(\vec{x})}{N_2(\vec{x})}\right),
\label{eq:Beta}
\end{equation}
with  $N_1$ and $N_2$ being  the populations in the lower and upper states of the lasing transition.}
Under these conditions, one can  apply the  FDT to find  the correlations of the noise~\cite{Eckhardt1982}: 
\begin{align}
\left<\tilde{\vec{F}}_S(\vec{x},\omega)
\tilde{\vec{F}}\textsuperscript{*}\hspace{-5pt}_S(\vec{x}',\omega')\right>=
4\hbar\omega^4\mathrm{Im}\left[\varepsilon(\vec{x},\omega)\right]
\coth\left(\frac{\hbar\omega\beta(\vec{x},\omega)}{2}\right)\delta(\vec{x}-\vec{x}')\delta(\omega-\omega').
\label{eq:FDT}
\end{align}
{ Here, $\varepsilon(\vec{x},\omega)$  is the dispersive permittivity of the laser, which includes nonlinear gain saturation above the lasing threshold [$\varepsilon$ is defined in Table 1 and by  the square brackets in \eqref{steady ME}]. The inverse temperature, $\beta$,   and the imaginary part of the permittivity, $\mathrm{Im}[\varepsilon]$, are  negative in gain regions (where the inversion $D \equiv N_2-N_1$ is positive) while both are positive elsewhere.  } In our approach (and also in ~\citeasnouns{Henry1986,Duan1990,Exter1995}), $\vec{F}_S$ represents the \emph{fluctuating spontaneous emission field}.    An equivalent description of laser noise can be obtained by introducing \emph{fluctuating currents in the atomic variables}  [\eqref{p-dot} and \eqref{d-dot}], instead of  $\vec{F}_S$, but we showed in \citeasnoun{cerjan2015quantitative} that the formulations are equivalent.

A recent advance in the  theory of microstructured lasers~\cite{Tureci2006,Tureci2007,Ge2010} shows that in many cases,  the Maxwell--Bloch equations can be greatly simplified: The inversion in most  microlasers is nearly stationary\footnote{Since micro-structured lasers have a large free spectral range (i.e., the mode spacing scales as $~1/L$, where $L$ is the length-scale of the structure), the beating terms in \eqref{d-dot} can be neglected~\cite{Tureci2006}.} and, therefore,  there exists a stable steady-state solution of  the form
\begin{equation}
\vec{E}(\vec{x},t)=\sum_\mu \vec{E}_\mu(\vec{x})a_{\mu0} e^{-i\omega_\mu t}.
\label{eq:ss-solution}
\end{equation}
The Maxwell--Bloch equations can be reduced to a single  Maxwell-type equation of the form
\begin{equation}
\left(\nabla\times\nabla\times -\omega_\mu^2\left[
\varepsilon_c(\vec{x},\omega)+
\frac{\gamma_\perp}{\omega_\mu-\omega_a+i\gamma_\perp}
\frac{D_0\,F(\vec{x})}{1+
\sum_\nu 
\frac{\gamma_\perp^2}{(\omega_\nu-\omega_a)^2+\gamma_\perp^2}|a_{\nu0}|^2|\vec{E}_\nu(\vec{x})|^2}
\right]\right) \vec{E}_\mu(\vec{x})=0.
\label{eq:steady ME}
\end{equation}
This is a dispersive  nonlinear eigenvalue problem, whose solutions determine the steady-state lasing frequencies $\omega_\mu$, amplitudes $a_{\mu0}$, and modes $\vec{E}_\mu(\vec{x})$, which can be found by employing  numerical algorithms (as outlined in~\citeasnoun{Esterhazy2013}).  The set of assumptions underlying the derivation of \eqref{steady ME} are commonly abbreviated as SALT---the steady-state \emph{ab-initio} laser theory.

When noise is introduced, the laser field can still be  approximated by \eqref{ss-solution}, but now the complex amplitudes, $a_\mu(t)$, vary  over  time. In~\citeasnoun{pick2015ab}, we derive  dynamical equations for $a_\mu(t)$ by using \emph{numerical solutions} of the SALT equation [\eqref{steady ME}] while  treating the effect of noise \emph{analytically}. 
{ A weak noise  causes small intensity fluctuations relative to the steady-state  intensity i.e., $|a_\mu(t)|^2\approx|a_{\mu0}|^2$ (this assumption  breaks down near the lasing threshold).}
In the single-mode regime,  we find 
\begin{align}
\dot{a}_\mu(t) = \int \!d\vec{x} \,c_{\mu\mu}(\vec{x}) \gamma(\vec{x})\int^t\!\!
dt'e^{-\gamma(\vec{x})(t-t')}\left(a_{\mu0}^2 - |a_\mu(t')|^2\right)a_\mu(t) + f_\mu(t),
\label{eq:Osci-single}
\end{align}
 \begin{table}[t!]
  \center
{\footnotesize
\begin{tabular}{c c c }
Quantity    & Symbol   & Definition \\[0.1ex] \hline\hline \\[-1ex]
\begin{tabular}{@{}c@{}} SALT permittivity  \end{tabular}
& $\varepsilon(\vec{x},\omega)$&   
$\varepsilon_c(\vec{x},\omega)+
\frac{\gamma_\perp D_0F(\vec{x})}{\omega-\omega_a+i\gamma_\perp}
\left[1+\displaystyle\sum_\mu
\frac{\gamma_\perp^2}{(\omega_\mu-\omega_a)^2+\gamma_\perp^2}|a_{\mu0}|^2|\vec{E}_\mu|^2\right]^{-1}$\\[2ex]
\begin{tabular}{@{}c@{}} Nonlinear restoring force\end{tabular}
&$c_{\mu\nu}(\vec{x})$  & $\frac{-i\omega_{\mu}^2
\tfrac{\partial\varepsilon(\vec{x},\omega_\mu)}{\partial |a_{
\nu0}|^2} \vec{E}_\mu^2(\vec{x})}
{\Scale[0.9]{\displaystyle\int\!dx\,\left.\tfrac{d}{d\omega}
\left[\omega^2\varepsilon(\vec{x},\omega)\right]\right|_{\omega_\mu}\vec{E}_\mu^2(\vec{x})}}$ \\[4ex]
\begin{tabular}{@{}c@{}} Dressed decay rate\end{tabular}
& $\gamma(\vec{x})$ &  $
\gamma_\parallel 
\left(1+\displaystyle\sum_\mu
\frac{\gamma_\perp^2}{(\omega_\mu-\omega_\mathrm{a})^2+\gamma_\perp^2}
\hspace{2pt}|a_{\mu0}|^2|\vec{E}_\mu(\vec{x})|^2\right).$ \\[4ex]
\begin{tabular}{@{}c@{}} Noise amplitude  \end{tabular}
&   $R_{\mu\nu}(\omega)$ & $
2\hbar\omega_\mu^4\hspace{2pt}
\frac{\Scale[0.9]{\displaystyle\int\!\! dx\hspace{2pt} {|\vec{E}_\mu(\vec{x})|}^2\mbox{Im}\hspace{2pt}\varepsilon(\vec{x},\omega)
\coth\left(\frac{\hbar\omega_\mu\beta(\vec{x},\omega)}{2}\right) }}{
{\left|\Scale[0.9]{\displaystyle\int\!dx\,\left.\tfrac{d}{d\omega}
\left[\omega^2\varepsilon(\vec{x},\omega)\right]\right|_{\omega_\mu}\vec{E}_\mu^2(\vec{x})} \right|}^2
}\cdot\delta_{\mu\nu}$\\[4ex]
\end{tabular}}
\caption{The coefficients of the single- and multi-mode generalized rate    equations [\eqref{Osci-single} and \eqref{Multimode-TCMT}], {  expressed in terms of the  laser parameters [cavity permittivity, $\varepsilon_c(x)$,  gain frequency and bandwidth, $\omega_a$ and $\gamma_\perp$, and  pump intensity and spatial profile,  $D_0$ and $F(x)$] as well as the laser  steady-state properties [SALT frequencies $\omega_\mu$, mode amplitudes, $a_{\mu0}$, and mode profiles, $\vec{E}_\mu(\vec{x})$].}  The definitions  are borrowed from \citeasnoun{pick2015ab}.}
\label{table:Factors2}
\end{table}
where    the parameters $c_{\mu\mu}(\vec{x})$, $\gamma(\vec{x})$, and $a_{\mu0}$  are   obtained from SALT (as shown in  Table~1)~\cite{pick2015ab}.
The nonlinear restoring force, $c_{\mu\mu}(\vec{x})$, can be thought of as an effective gain rate (being proportional to the product of the lasing frequency $\omega_\mu$ and  pump amplitude $D_0$). The dressed relaxation rate, $\gamma(\vec{x})$, is a sum of  the bare atomic-relaxation  rate, $\gamma_\parallel$, and a nonlinear spatially-inhomogeneous  term, which turns on at the lasing threshold.  
Last, the noise is represented by a random Langevin term, $f_\mu(t)$, and only its amplitude $R_{\mu\mu}$   [defined via  $\langle f_\mu(t)f_\mu^*(t')\rangle = R_{\mu\mu}\delta(t-t')$] determines the (ensemble-averaged) noise spectrum. 
{  Treating spontanteous emission as white noise~\cite{Henry1982} (i.e., uncorrelated in time)
is equivalent to assuming that the noise autocorrelation function [$R_{\mu\mu}(\omega)$] is nearly constant for frequencies within the lasing peaks. This assumption is valid when the lasing linewidths are much narrower than the gain bandwidth.   The effect of colored noise can be incorporated into our approach, as mentioned in \secref{discussion-sec}.}
A solution of  \eqref{Osci-single}  is shown in \figref{RO-illustration}(c) for a particular  realization of the noise process, $f_\mu(t)$, with  parameters $a_\mu(0) = 5, a_{\mu0} = 1, R_{\mu\mu} = 1.44\cdot10^{-4}\,s^{-1}, \int\!\!d\vec{x}\,c_{\mu\mu}(\vec{x})= 0.19+1.18i\,s^{-1}$ and a constant atomic-relaxation rate, $ \gamma(\vec{x}) = 0.0025\,s^{-1}$ (which is a good approximation  near  threshold, because the nonlinear inhomogeneous term is much smaller than the bare rate).  These parameters correspond to a   type-B laser (i.e., with comparable atomic and light relaxation rates) and, indeed,  the solution  reveals RO dynamics. In \citeasnoun{pick2015ab}, we used  \eqref{Osci-single} to compute the central-peak linewidths. In this work, we use it to compute the entire noise spectrum,  as shown in the next section.

\section{The noise spectrum of single-mode lasers\label{Single-result-sec}}
\subsection{Formula for the noise spectrum}

Before diving into the details of the derivation of the single-mode formula  (in \secref{derivation-text}), we summarize our  results: the new formula, its validation, and its consequences.
The  noise spectrum of a single-mode laser with lasing-frequency $\omega_\mu$ is
\begin{align}
&S_\mu(\omega) = 
\underbrace{\frac{{\Gamma_0(\omega_\mu)}(\alpha_1^2+1)}{(\omega-\omega_\mu)^2 + \left[\tfrac{{\Gamma_0(\omega_\mu)}}{2}(\alpha_1^2+1)\right]^2}
\left(1 - \frac{{\Gamma_0(\omega_\mu)}(\alpha_2^2+1)}{4\Gamma} \right)}_
{\text{central peak}}+\nonumber\\
&\underbrace{\frac{{\Gamma_0(\omega_\mu-\Omega)}(\alpha_2^2+1)/4}{\Gamma_\mathrm{SB}^2+(\omega-\omega_\mu+\Omega)^2}
\left(1 + \frac{4\alpha_3}{\alpha_2^2+1}\cdot
\frac{\Gamma}{\Omega}+\frac{3\alpha_2^2-1}{\alpha_2^2+1}\cdot
\frac{\omega-\omega_\mu+\Omega}{\Omega}\right)}_
{\text{red  sideband}}+\nonumber\\
&\underbrace{
\frac{{\Gamma_0(\omega_\mu+\Omega)}(\alpha_2^2+1)/4}{\Gamma_\mathrm{SB}^2+(\omega-\omega_\mu-\Omega)^2}
\left(1 - \frac{4\alpha_3}{\alpha_2^2+1}\cdot\frac{\Gamma}{\Omega}-
\frac{3\alpha_2^2-1}{\alpha_2^2+1}\cdot
\frac{\omega-\omega_\mu-\Omega}{\Omega}\right)}_
{\text{blue sideband}}.
\label{eq:single-mode-result}
\end{align}

  \begin{table}[t!]
 \center
{\footnotesize
\begin{tabular}{c c c }
Quantity    & Symbol   & Definition \\[0.1ex] \hline\hline 
\begin{tabular}{@{}c@{}c@{}}Phase diffusion coefficient\end{tabular}
& $\Gamma_0(\omega)$&  $  R_{\mu\mu}(\omega)/2a_{\mu0}^2$ \\[2ex]
\begin{tabular}{@{}c@{}c@{}}RO frequency\end{tabular}
&$\Omega$  
&   
$ \left[2a_0^2\int \!d\vec{x}\, \mathrm{Re}\,c_{\mu\mu}(\vec{x})\gamma(\vec{x})\right]^{1/2}$\\[2ex]
\begin{tabular}{@{}c@{}c@{}}RO decay rate\end{tabular}
&$\Gamma$  &
$ \int \!d\vec{x}\, \gamma(\vec{x})/2 $\\[2ex]
\begin{tabular}{@{}c@{}c@{}}Sideband linewidth\end{tabular}
&$\Gamma_\mathrm{SB}$  
&   
$ \Gamma_0(\alpha_1^2+1)+\Gamma$\\[2ex]
\begin{tabular}{@{}c@{}c@{}}Linewidth enhancement\end{tabular}
&$\alpha_1$  
&   $ \frac{\int\! d\vec{x}\,\mathrm{Im}\,[c_{\mu\mu}(\vec{x})]}{\int\! d\vec{x}\,\mathrm{Re}\,[c_{\mu\mu}(\vec{x})] }$\\[2ex]
\begin{tabular}{@{}c@{}c@{}}Sideband power fraction\end{tabular}
&$\alpha_2$  &   $ \frac{\int \!d\vec{x}\,\gamma(\vec{x})\mathrm{Im}\,[c_{\mu\mu}(\vec{x})]}{\int\! d\vec{x}\,\gamma(\vec{x})\mathrm{Re}\,[c_{\mu\mu}(\vec{x})] }$\\[2ex]
\begin{tabular}{@{}c@{}c@{}}Asymmetry factor\end{tabular}
&$\alpha_3$  &   $ \frac{\int \!d\vec{x}\,\gamma(\vec{x})^2\mathrm{Im}\,[c_{\mu\mu}(\vec{x})]}{
\left[\int \!d\vec{x}\,\gamma(\vec{x})\mathrm{Re}\,[c_{\mu\mu}(\vec{x})] \right]
\left[\int\!d\vec{x}\,\gamma(\vec{x})\right]}$\\[2ex]
\end{tabular}}
\caption{Coefficients of the single-mode noise spectrum [\eqref{single-mode-result}],  {  expressed  in terms of quantities obtained from SALT: the steady-state modal amplitudes, $a_{\mu0}$, and the oscillator-equation coefficients, $c_{\mu\mu}(\vec{x})$,  $\gamma(\vec{x})$, and $R_{\mu\mu}$,  defined in Table 1.}}
\label{table:Factors1}
  \end{table}
The first term  corresponds to the central Lorentzian peak, while the second and third terms are the red  and blue  RO sidepeaks.   In Table 2, we express all the parameters  from   \eqref{single-mode-result}  in terms of  the coefficients of the generalized rate equation [\eqref{Osci-single}]. {  For ease of notation, we omit the subscript $\mu$ from the coefficients. }Since these coefficients  are functions of the SALT solutions (as shown in Table 1), the evaluation of \eqref{single-mode-result} requires  no additional  free parameters besides those appearing in the Maxwell--Bloch equations [\eqref{Maxwell-Bloch}].  The central peak is centered around the SALT lasing frequency, $\omega_\mu$, and its   linewidth is  the product of the phase-diffusion coefficient, ${\Gamma_0}(\omega_\mu)$,    and the amplitude--phase-coupling  enhancement factor,   $\alpha_1^2+1$. 
 Since some of the noise power goes into the sidepeaks,  the amplitude of the central peak is reduced by a factor of  $1-\tfrac{\Gamma_0(\omega_\mu)}{4\Gamma}(1+\alpha_2^2)$, where   $\Gamma$ is the rate at which ROs decay and $\alpha_2$ is the second generalized phase--amplitude-coupling factor.   
 The RO sidepeaks are Lorentzians, whose  center-frequency and linewidth are $\omega_\mu\pm\Omega$ and    $\Gamma_\mathrm{SB}$ respectively.   The amplitude of the blue  and lred sidepeaks differs by a factor of $\frac{4\alpha_3}{\alpha_2^2+1}$, where $\alpha_3$ is the third generalized amplitude--phase-coupling factor.

 Our new formula  [\eqref{single-mode-result}]  is formally similar to the result of~\citeasnoun{Exter1992}, but here  we  obtain three kinds  of generalized  $\alpha$ factors, while in \citeasnoun{Exter1992} they are the same. {  In  \citeasnoun{Exter1992}, the $\alpha$ factoer is given by the traditional expression $\alpha_{1,2,3} =\frac{\mathrm{Re}[\Delta n]}{\mathrm{Im}[\Delta n]}$,  where $\Delta n$ is the change in index of refraction following a noise-driven perturbation~\cite{Henry1982}. 
 In contrast, our  generalized $\alpha$ factors  are spatial averages of the  refractive index change  with different weight factors (as defined in Table 2 and discussed in Sec.~3.2). }  While the parameters in our formula  are obtained directly from the Maxwell--Bloch equations,  the parameters in \citeasnoun{Exter1992}  are  expressed  in terms of many additional  parameters (such as the mode volume,  confinement factor, cold-cavity decay rate, effective differential gain, gain saturation coefficient, etc.) and, quantitatively, can   only be obtained by empirical fits. Similar to previous work, our derivation of \eqref{single-mode-result} assumes that  $\Gamma\ll\Omega$, which implies that the sidepeaks have little overlap with the central lasing peak. 

\begin{figure}[t]
\centering
                 \includegraphics[width=1\textwidth]{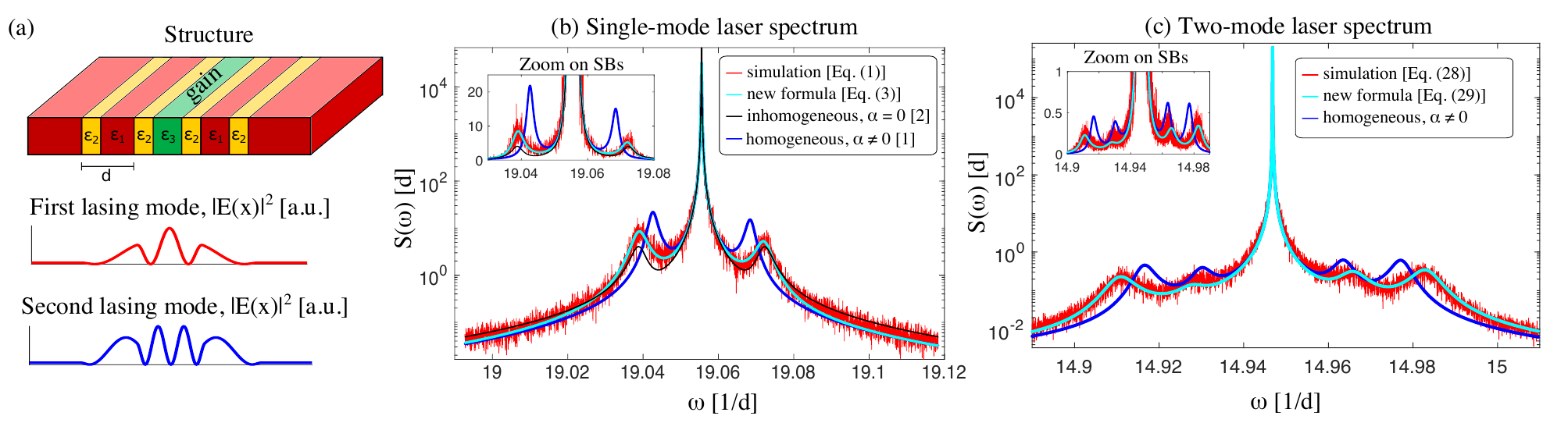}

                                    \caption{
(a) Top: A periodic stack of layers with alternating permitivities ($\varepsilon_1, \varepsilon_2$) and thicknesses ($d_1, d_2$), with a  defect layer (with $\varepsilon_3$ and $d_3$). 
The parameters (see text) are chosen such that  the structure has two cold-cavity localized modes inside the band gap.  Gain is added in the three central layers in order to make the gap modes lase. Bottom: Intensity profiles of  the  first and second  lasing modes (with threshold frequencies $\omega_1 = 19.05$ and $\omega_2 = 14.95$ respectively).
 (b) Spectrum of a single-mode laser, on a log-linear scale, computed by time-stepping \eqref{Osci-single} (red) and by evaluating our single-mode formula~[\eqref{single-mode-result}] (cyan) and earlier results:~\cite{Exter1992} (black) which neglected $\alpha-$factor corrections and~\cite{pick2015ab} (blue) which neglected inhomogeneity and nonlinearity of the modes and  gain.  Inset: Magnification of  the sidepeaks, plotted on a linear scale, which shows the asymmetry of the peaks. (c)  Spectrum of a   multimode laser. We compare the numerical solution of the stochastic equations [\eqref{Multimode-TCMT}] (red) with our multimode formula [\eqref{Multimode-spectrum}]  (cyan). Additionally, we plot the homogeneous limit of our formula (black) . Inset: Zoom on the sidepeaks.}
\label{fig:numerical-TCMT}
  \end{figure}

\subsection{Validation and main predictions of the formula }
We validate our single-mode formula [\eqref{single-mode-result}] by comparing it  with brute-force simulations of the generalized rate  equations [\eqref{Osci-single}] and with previous theories~\cite{pick2015ab,Exter1992} (\figref{numerical-TCMT}).  Since we expect~\eqref{single-mode-result} to deviate from the traditional results  in the limit of substantially different $\alpha$ factors, we study a numerical example where the $\alpha$ factor can be easily tuned: A  periodic array of  dielectric slabs with a defect at the center of the structure and     gain in the defect area (we discussed a similar structure in \citeasnoun{pick2015ab}). Our motivation to study this structure is the fact that the traditional  $\alpha$ factor is proportional to the detuning of the gain resonance  from the lasing frequency~\cite{Lax1966}; since the frequency of the defect mode is unaltered by small changes in the gain, one can vary $\alpha$ by varying the resonance  of the gain\footnote{{ A possible candidate system for measuring this effect  is a  Zeeman-split laser~\cite{shelton1992phase}, where   the frequency of the lasing transition varies in proportion to an external magnetic field.}}.  The structure is shown in panel (a).  The parameters are
 $\varepsilon_1 = 1, \varepsilon_2 = 16, \varepsilon_3 = 7, d_1 = 0.2a, d_2 =\tfrac{\sqrt{\varepsilon_1}a}{\sqrt{\varepsilon_1}+\sqrt{\varepsilon_2}} = 0.8a, d_3 = 0.2a, \gamma_\parallel = 0.006, \omega_a = 18$ and  $\gamma_\perp = 1$ in (b) [and  $\omega_a = 17$ and  $\gamma_\perp = 2$ in (c)]. Here,  $a$  is the unit-cell size and the frequency unit is $2\pi c/a$.     We employ a finite-difference frequency-domain (FDFD)~\cite{christ1987three,champagne2001fdfd} approach to   discretize the SALT equations,  and use the algorithm from~\citeasnoun{Esterhazy2013} to  obtain  the steady-state  modes [$\vec{E}_\mu(\vec{x})$],  frequencies ($\omega_\mu$), and amplitudes ($a_{\mu0}$).  Using these  solutions, we compute the coefficients from Table 1, which we use both to evaluate our spectral  formula [\eqref{single-mode-result}] and as the starting point for numerical simulations of \eqref{Osci-single}. { 
 The simulations include time-stepping of  \eqref{Osci-single} (by implementing a standard  Euler scheme for   stochastic ordinary differential equations~\cite{Press2007}) and taking  the ensemble  average of  the  Fourier transform of the mode intensity $|a_\mu|^2$ (also called  the periodogram of the signal~\cite{oppenheim1999discrete}). }

The results are shown in panel (b). An important advantage of the  new formulation  is that it  correctly accounts for  the spatially dependent enhancement of the atomic relaxation rate, $\gamma(\vec{x})$,  above the lasing threshold  (defined in Table 1). This enhancement  affects the sideband spectrum since both the oscillation frequency and sideband linewidth depend on $\gamma(\vec{x})$ (see Table 2). Previous treatments, which assumed either that the relaxation rate is independent of the field~\cite{vahala1983semiclassicalB} or that it is  constant (fixed at the unsaturated value)~\cite{Exter1992}, underestimated the broadening and shifting of the sidepeaks.  In Figs.~2(b--c),  we demonstrate that  our formula (cyan)
matches the numerically simulated noise spectrum (red), while  homogeneous models, which corresponds to assuming a  bare relaxation  rate (black) or an   unsaturated rate (blue) fail.

\begin{figure}[t]
\centering
                 \includegraphics[width=0.8\textwidth]{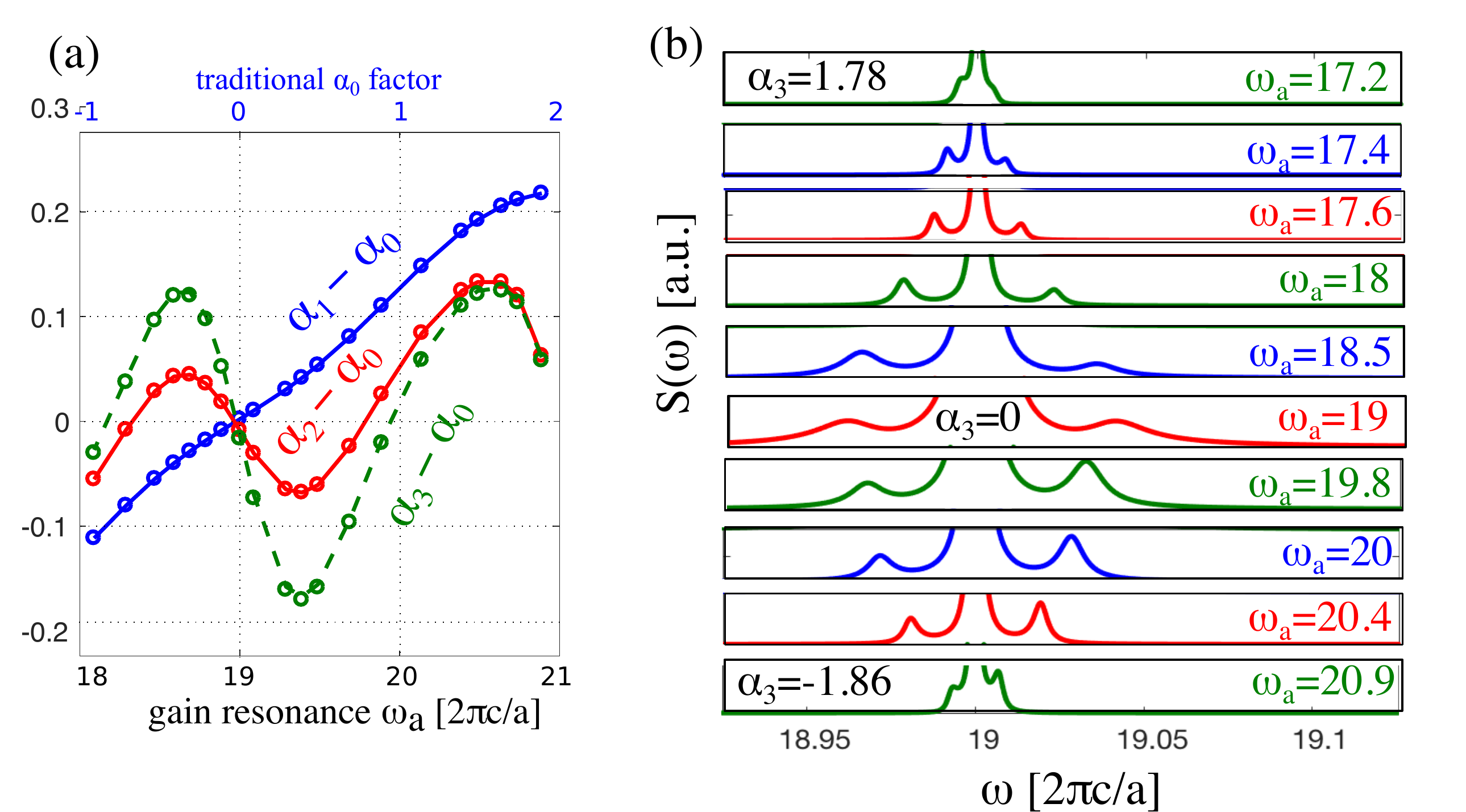}

\caption{(a) Deviation of the generalized $\alpha$ factors ($\alpha_{1,2,3}$ in Table 2) from the traditional factor ($\alpha_0=\tfrac{\omega_a-\omega_\mu}{\gamma_\perp}$) for the structure from \figref{numerical-TCMT}a. 
The plot shows the $\alpha$ factors at a fixed pump power for varying gain frequencies. Large deviations are evident for large detunings. Most notably, $\alpha_3$ deviates non-monotonically from $\alpha_0$ in the shown frequency range.
(b) Sideband spectrum for gain-resonance frequencies in the range $\omega_a  = 17.2\hdots21$. When $\alpha_3>0$, the red  sidepeak are stronger than the blue  sidepeak, and this picture is reversed when $\alpha_3<0$. 
}
\label{fig:RO-properties}
  \end{figure}

\figrefbegin{RO-properties}(a) presents a comparison of  the traditional and   generalized amplitude--phase coupling factors\footnote{{ 
A comparison between the traditional and new $\alpha$ factors can be made by using the definitions in Table 2,   which relate the generalized $\alpha$ factors to the nonlinear coefficient $c_{\mu\mu}$, and Table 1, which  defines $c_{\mu\mu}$ in terms of the derivative of the permittivity, $\varepsilon$. The permittivity and the index are related via $\varepsilon = n^2$ for nonmagnetic media (where $\mu = 1$) (see \citeasnoun{pick2015ab} for details).}}.
The  traditional $\alpha$ factor  was    introduced by Lax~\cite{Lax1966}, where he used a zero-dimensional model (which neglects inhomogeneity in the pump and the fields) to explain central-peak linewidth broadening in detuned-gas lasers. \citeasnoun{Lax1966} shows  that  the amplitude--phase coupling  is equal to the detuning of the lasing frequency from the atomic resonance, i.e., $\alpha_0=\frac{\omega_0-\omega_\mathrm{a}}{\gamma_\perp}$. 
Later work by Henry~\cite{Henry1982} found  that in semiconductor lasers, the amplitude-phase coupling is $\tilde{\alpha}_0=\frac{\mathrm{Re}[\Delta n]}{\mathrm{Im}[\Delta n]}$,  where $\Delta n$ is the change in index of refraction following a noise-driven perturbation.   In \citeasnoun{pick2015ab}, we  showed that the Lax and Henry definitions are equivalent and  that, more generally, the amplitude-phase coupling ($\alpha_1$) is given by the  ratio of the  \emph{spatial averages} of the real- and imaginary-index fluctuations  (see definition in Table 2).
Moreover, we  showed  that the difference  between the traditional and the generalized  factors, $\alpha_1-\alpha_0$,  increases with increasing $\alpha_0$. Motivated by this prediction, we present  in \figref{RO-properties}(a) the deviation  of the   generalized  $\alpha$ factors ($\alpha_1, \alpha_2, \alpha_3$) from  the traditional $\alpha_0$ as a function of gain-center frequency  $\omega_a$.
 We find that all three factors  deviate substantially from $\alpha_0$ at large detunings.  All the data points in the plot are obtained at a fixed pump power ($D_0 = 0.095$).  The  relaxation rates of the inversion and polarization  are $\gamma_\parallel = 0.006$ and $\gamma_\perp = 1$, as in  \figref{numerical-TCMT}(b).   
 
{ \figrefbegin{RO-properties}(b) demonstrates   the dependence  the sideband asymmetry on the generalized  factor $\alpha_3$.  We compute the entire noise  spectrum for several gain-center frequencies  in the range $\omega_a - \omega_\mu\in(-1.8,2)$, with $\gamma_\parallel = 0.02$, $\gamma_\perp = 1$ and  $D_0 = 0.095$. 
From \eqref{single-mode-result}, one can see that  the asymmetry is controlled by $\alpha_3$. In this numerical example, $\alpha_0\approx1$ and $\alpha_3$ differs from $\alpha_0$ by approximately $10\%$ [see  \figref{RO-properties}(a)].  The traditional factor $\alpha_0$ changes sign when the gain frequency is equal to   the lasing frequency, so we expect the asymmetry of the sidebands to change sign as we sweep the gain center frequency across the cavity resonance. This trend is evident in \figref{RO-properties}(b).  Since  $\alpha_3$ changes sign in the range $\omega_0-\omega_\mathrm{a}\in(0,1)$,  the red  sidepeaks are weaker  than the blue sidepeaks, in contrast to the more common case of positive-$\alpha$ semiconductor lasers~\cite{Osinski1987}, where red sidebands are stronger.}

\subsection{Derivation outline\label{derivation-text}}
 In this section, we   outline the derivation of \eqref{single-mode-result}, leaving the  detailed explanations to  appendix A. 
Our derivation is inspired by the approach of \citeasnoun{Exter1992}, but since we use the \emph{ab-initio} dynamical oscillator equations [\eqref{Osci-single}] instead of the traditional laser rate equations, our derivation is  more involved and the results are more general. Our starting point is the Wiener--Khintchine theorem~\cite{kittel2004elementary}, which relates the laser-noise spectrum to the Fourier transform of the autocorrelation function $\left<a(t)a^*(0)\right>$ [where angle brackets  denote an  ensemble average over realizations of the noise process].  Since intensity and phase fluctuations
have distinct roles in determining the noise spectrum (as explained in the introduction), it is convenient to  write the complex mode amplitude, $a$,  in the form~\cite{Lax1967a}:
\begin{equation}
a(t) =   a_0  e^{-u(t)+i\phi(t)},
\label{eq:ansatz}
\end{equation}
The autocorrelation of $a$ can be written as:
\begin{gather}
\frac{\langle a(t)a^*(0)\rangle}
{\langle|a(0)|^2\rangle} =\langle\exp\left\{-\left[u(t)+u(0)\right]+i\left[\phi(t)-\phi(0)\right]\}\right\}\rangle /\langle\exp{[-2\langle u(0)\rangle]}
\rangle \approx\nonumber\\
\exp \Scale[1.5]{\{ }
-
\underbrace{\tfrac{\langle[\phi(t)-\phi(0)]^2\rangle}{2}}_{\text{phase variance}}
\,+\!\underbrace{\tfrac{\langle[u(t)+u(0)]^2\rangle}{2}\rangle}_{\text{intensity correlations}}
-i\underbrace{\langle [u(t)+u(0)][\phi(t)-\phi(0)]\rangle}_{\text{cross term}}\Scale[1.5]{\}}/\langle\exp{[-2\langle u(0)\rangle]}.
\label{eq:autocorrelation}
\end{gather}
The approximation in going from the first to second line can be justified as follows:
First, we expand   the exponent in   a Taylor series.   Since intensity fluctuations  are smaller than the steady-state intensity, all the terms involving $u$ are small and we  keep  only the  leading-order  terms in the expansion.   The phase variance [i.e., the $\phi^2$ term, given explicitly in \eqref{PhiVarResult} below]  is the sum of a ``Brownian drift'' term that grows linearly with time and a  small RO term.  The  phase drift  is the result of a Wiener (Brownian-motion) process of many uncorrelated spontaneous-emission ``kicks'' and, from the central-limit theorem~~\cite{Feller1945,Feller1968}, it follows that  it is a Gaussian variable.  The RO term is small and  we keep only the corresponding leading term in the expansion. With these assumptions,  we can move the ensemble average from the second equality on the  first line  inside the exponent and obtain the second line. This step  is exact for log-normal distributions~\cite{johnson1994continuous} (i.e., the exponent of a Gaussian phase), while it is a very good approximation for small fluctuations. Previous authors used a similar identity~\cite{Exter1995,cohen1989spectral}, but incorrectly justified it by saying that all the  variables are Gaussian, while clearly $u$ and $\phi$ are not Gaussian because they perform relaxation oscillations.

In order to relate the autocorrelation, $\langle a^*(t)a(0)\rangle$, to the steady-state laser properties, we  need to obtain  explicit expressions for    the second-order moments: the phase variance, the intensity autocorrelation, and the cross term, defined in \eqref{autocorrelation}. To this end, we  substitute \eqref{ansatz} into \eqref{Osci-single} and linearize the resulting expression by assuming that intensity fluctuations  are small compared to the steady-state intensity  (i.e., $|u|\ll1$). (Note that by linearizing the equations, we lose the  higher-order RO peaks, but obtain accurate formulas for the first-order sidepeaks.)  This procedure yields 
\begin{subequations}
\begin{gather}
\dot{\phi}(t)= \int\! d\vec{x}\, B(\vec{x})\xi(\vec{x},t) +f_I(t)/a_0 \,,
\label{eq:linearized1}
\\
\dot{u}(t)= -\int\! d\vec{x}\,  A(\vec{x}) \xi(\vec{x},t) + f_R(t) / a_0 \,,
\label{eq:linearized2}\\
\dot{\xi}(\vec{x},t)=-\gamma(\vec{x})\xi(\vec{x},t)+\gamma(\vec{x}) u(t) \,,
\label{eq:linearized3}
\end{gather}
\label{eq:linearized}
\end{subequations}
\hspace{-3pt}where we introduced the time-delayed  intensity fluctuation,
$\xi(\vec{x},t)\equiv\int^t\!\!dt'e^{-\gamma(\vec{x})(t-t')}u(t')$, in order to turn the  integro-differential equations   into a set of ordinary-differential equations (ODEs)~\cite{pick2015ab}. 
We also introduced  $A(\vec{x})$ and $B(\vec{x})$ to denote   the real and imaginary parts of the nonlinear restoring force $2a_0^2c(\vec{x})$,  and  $f_{R}(t)$ and $f_{I}(t)$ are   the real and imaginary parts of the Langevin noise term. 
We proceed by taking the Fourier transform  of the linearized equations [\eqref{linearized}]. 
We solve the frequency-domain equations and obtain
\begin{subequations}
\begin{gather}
\tilde{u}(\omega)
=\frac{1}{i\omega + 
\int\!\!d\vec{x} \frac{A(\vec{x})\gamma(\vec{x})}{\gamma(\vec{x})+ i\omega}}\cdot
\frac{\tilde{f}_R(\omega)}{a_0}
\\
\tilde{\xi}(\vec{x},\omega) =\frac{\gamma(\vec{x})}{\gamma(\vec{x}) + i\omega}
\cdot\frac{1}{i\omega + \int\!\!d\vec{x} \frac{A(\vec{x})\gamma(\vec{x})}{\gamma(\vec{x})+ i\omega}}\cdot
\frac{\tilde{f}_R(\omega)}{a_0}
\label{eq:PhiUxi}\\
\tilde{\phi}(\omega) =
\frac{\int\!\!dx
\tfrac{\gamma(x)B(x)}{\gamma(x) + i\omega}\cdot
}{i\omega + \int\!\!dx \frac{A(x)\gamma(x)}{\gamma(x)+ i\omega}}\cdot
\frac{\tilde{f}_R(\omega)}{i\omega a_0}
+
\frac{\tilde{f}_I(\omega)}{i\omega a_0},
\label{eq:PhiTilde}
\end{gather}
\label{eq:linearSOLS}
\end{subequations}

 As shown in  appendix~A,  the time-dependent second-order moments can be written in terms of integrals over the power spectral densities~\cite{Henry1986} 
 \begin{subequations}
\begin{gather}
\left<[\phi(t)-\phi(0)]^2 \right>=
\frac{1}{2\pi}
\iint_{-\infty}^\infty
{d}\omega{d}\omega' \,
\left<\tilde{\phi}(\omega)\tilde{\phi}^*(\omega')\right>
(1-e^{i\omega t})(1-e^{-i\omega' t}),
\label{eq:PhiVarFT}
\\
\left<[u(t) + u(0)]^2\right>  =
\frac{1}{2\pi}
\iint_{-\infty}^\infty
{d}\omega{d}\omega' \,
\left<\tilde{u}(\omega)\tilde{u}^*(\omega')\right>
(1+e^{i\omega t})(1+e^{-i\omega' t}),
\label{eq:uVar}
\\
\left<[\phi(t)-\phi(0)][u(t) + u(0)]\right> = 
\frac{1}{2\pi}
\iint_{-\infty}^\infty
{d}\omega{d}\omega' \,
\left<\tilde{\phi}(\omega)\tilde{u}^*(\omega')\right>
(1-e^{i\omega t})(1+e^{-i\omega' t}).  
\label{eq:CrossTerm}
\end{gather}
\label{eq:VARS}
\end{subequations}
\hspace{-2pt}Since the  integrands are meromorphic functions, these integrals can be computed by invoking the Cauchy residue theorem~\cite{Arfken2006},  which relates the integrals to the residues and poles of the integrands.  The pole of $\tilde{\phi}$ at $\omega = 0$ produces the central-peak linewidth, which we computed in \citeasnoun{pick2015ab}. 
 In order to see the remaining poles more clearly, we introduce the  approximation:
\begin{align}
\frac{1}{i\omega + \int\!\!d\vec{x} \frac{A(\vec{x})\gamma(\vec{x})}{\gamma(\vec{x})+ i\omega}} =
\frac{1}{\int\!\!d\vec{x} 
\frac{\left[i\omega(\gamma(\vec{x})+ i\omega)+A(\vec{x})\gamma(\vec{x})\right]}{\gamma(\vec{\vec{x}})+ i\omega}
}  \approx
\frac{i\omega}{\int\!\!d\vec{x} \left[i\omega (\gamma(\vec{x})+i\omega)+A(\vec{x})\gamma(\vec{x})\right]} .
\label{eq:APPROX}
\end{align}
In the last equality,  we assumed  that $\gamma(\vec{x}) + i\omega\approx i\omega$ for all $\vec{x}$, which holds near the RO frequencies in the limit of resolved sidepeaks, that is, for $\omega\approx\Omega \gg\Gamma$, using the definitions
\begin{equation}
\int\!\!d\vec{x} \left[i\omega (\gamma(\vec{x})+i\omega)+A(\vec{x})\gamma(\vec{x})\right] = 
-\omega^2 + i\omega  \Scale[0.9]{\left[\int\!\!d\vec{x}  \,\gamma(\vec{x})\right]}
+  \Scale[0.9]{\left[\int\!\!d\vec{x}A(\vec{x})\gamma(\vec{x})\right] }
\equiv -\omega^2+2i\omega\Gamma+\Omega^2 .
\label{eq:13}
\end{equation}
From \eqref{13}, one can see that the denominator of \eqref{APPROX} is a second-degree polynomial that vanishes at $\pm \Omega+i\Gamma$ (in the limit of $\Omega\gg\Gamma$). These zeros produce the RO sidepeaks in the noise spectrum. 
By collecting the results, we find
\begin{subequations}
\begin{gather}
\left<[\phi(t)-\phi(0)]^2 \right>= 
\tfrac{R_0}{a_0^2}\left(1+\alpha_1^2\right)t+ 
\tfrac{R_\pm\alpha_2^2}{2a_0^2\Gamma}
(1-e^{-\Gamma t}\cos\Omega t)  -
\tfrac{3R_\pm\alpha_2^2}{2a_0^2\Omega} e^{-\Gamma t}\sin\Omega t
\label{eq:PhiVarResult}
\\
\left<[u(t) + u(0)]^2\right>
=\frac{R_\pm}{2\Gamma a_0^2}
(1+\cos\Omega t e^{-\Gamma t})+
\frac{R_\pm}{2\Omega a_0^2}
\sin\Omega t e^{-\Gamma t}
\label{eq:uVarResult}
\\
\left<[\phi(t)-\phi(0)][u(t) + u(0)]\right>=  \tfrac{R_0\,\alpha_1}{a_0^2\,A}
+\tfrac{R_\pm\,\alpha_3}{a_0^2\Omega}\left(
-\tfrac{2\Gamma}{\Omega}
\cos\Omega t\,e^{-\Gamma t}+\sin\Omega t\,e^{-\Gamma t}\right),
\label{eq:CrossResult}
\end{gather}
\label{eq:moments-result}
\end{subequations}
{ \hspace{-3pt}where  $A \equiv \int\!dx\,A(\vec{x})$ and  all the parameters are defined in Table 2.  
We denote by $R_0$ and $R_\pm$ the autocorrelation evaluated at the lasing  and RO frequencies respectively, i.e.,  $R(\omega_\mu)$ and $R(\omega_\mu\pm\Omega)$.}
While  the phase variance [\eqref{PhiVarResult}] grows linearly in time,   the intensity autocorrelation and the cross term [\eqsref{uVarResult}{CrossResult}] do not {  show diffusive behavior,}  which is expected  because the nonlinear restoring force in the oscillator equations [\eqref{Osci-single}] prevents  intensity drift. 

After obtaining closed-form expressions for the second-order moments [\eqref{moments-result}], we substitute these results  into the autocorrelation [\eqref{autocorrelation}] and take the Fourier transform to obtain the noise spectrum.
The calculation can be simplified   when the central peak in the spectrum is much narrower than the sidebands [which holds when all the coefficients  in \eqref{moments-result} (i.e., $R_0(1+\alpha_1^2), R_\pm\alpha_2^2$, etc.) are much smaller than $\Gamma$]. In this regime, we can expand the exponentials in \eqref{autocorrelation} in a Taylor series around $R_\pm/\Gamma$ and obtain  \eqref{single-mode-result}.

\section{Noise spectrum of multimode lasers\label{MultiSec}}
We generalize our approach from  \secref{derivation-text} and obtain a formula for the multimode  noise spectrum. In this section, we present our result, and the derivation  details are given in  appendix B. The starting point of the derivation is the multimode dynamical equations for the complex  amplitudes $a_\mu$ [defined in \eqref{ss-solution}], which were derived in \citeasnoun{pick2015ab}:
\begin{align}
&\dot{a}_\mu(t) = 
\sum_{\nu  }\int\!d\vec{x}\, c_{\mu\nu}(\vec{x})  
\left[\gamma(\vec{x}) \int^t\!\!
dt'e^{-\gamma(\vec{x})(t-t')}\left(a_{\nu0}^2 - |a_\nu(t')|^2\right)\right]a_\mu(t) + f_\mu(t),
\label{eq:Multimode-TCMT}
\end{align}
where  $\mu,\nu=1\hdots M$, for $M$ lasing modes.  
In \citeasnoun{pick2015ab}, we used \eqref{Multimode-TCMT} to obtain the linewidths of the central lasing peaks. 
In appendix B, we complete the derivation of the multimode sidepeaks and find that the  Fourier transform of the autocorrelation $\langle a_\mu(t)a^*_\nu(t')\rangle$ is
\begin{align}
&
S_{\mu\nu}(\omega) = 
\underbrace{\frac{\Gamma_{\mu\nu}}{(\omega-\omega_\mu)^2+(\Gamma_{\mu\nu}/2)^2}
\left(1-\frac{
\sum_\sigma
{[\mat{S}^\sigma_{\mu\nu}+\mat{U}^\sigma_{\mu\nu}]}
}{2}\right)}_{\text{central peaks} } +\nonumber\\
&\underbrace{\Scale[0.9]{
\displaystyle\sum_\sigma 
\frac{\Gamma^{\mu\nu\sigma}_\mathrm{SB}}{(\omega-\omega_\mu+\Omega_\sigma)^2+{(\Gamma^{\mu\nu\sigma}_\mathrm{SB})}^2}
\left[\left(\frac{\mat{S}^\sigma_{\mu\nu}+\mat{U}^\sigma_{\mu\nu}+
2\mat{Y}^\sigma_{\mu\nu}}{2}\right)+
\frac{\Omega_\sigma+\omega-\omega_\mu}{\Gamma_\mathrm{SB}}
\left(\frac{
\mat{V}^\sigma_{\mu\nu}-\mat{T}^\sigma_{\mu\nu}+2\mat{X}^\sigma_{\mu\nu}
}{2}\right)
\right]}}_{\text{blue sidepeaks} } 
\nonumber\\
& \underbrace{\Scale[0.9]{
+\displaystyle\sum_\sigma
\frac{\Gamma^{\mu\nu\sigma}_\mathrm{SB}}{(\omega-\omega_\mu-\Omega_\sigma)^2+{(\Gamma^{\mu\nu\sigma}_\mathrm{SB})}^2}
\left[\left(\frac{\mat{S}^\sigma_{\mu\nu}+\mat{U}^\sigma_{\mu\nu}-
2\mat{Y}^\sigma_{\mu\nu}}{2}\right)-
\frac{\Omega_\sigma-\omega+\omega_\mu}{\Gamma_\mathrm{SB}}
\left(\frac{
\mat{V}^\sigma_{\mu\nu}-\mat{T}^\sigma_{\mu\nu}+2\mat{X}^\sigma_{\mu\nu}
}{2}\right)
\right]}}_{\text{red  sidepeaks} } .
\label{eq:Multimode-spectrum}
\end{align}  
\begin{table}[t]
\centering
{\footnotesize
\caption{Coefficients  of the multimode  formula [\eqref{Multimode-spectrum}].}
\begin{tabular}{c c  }
\arrayrulecolor{black}\hline\hline \\[-3ex]
  $A_{\mu\nu}(x) = 2a_{\mu0}a_{\nu0}\mathrm{Re}[c_{\mu\nu}(x)],\quad
  \mat{A} = \int\!\! dx \,\mat{A}(x)$&
   $ \mat{Q}_{+\sigma} = \displaystyle\sum_{\ell{m}{n}}
\frac{\mat{P}_{+\sigma}\mat{P}_{-{n}} \mat{R}_+\mat{P}_{-\ell}^\dagger\mat{P}_{+{m}}^\dagger}
{(\omega_{+\sigma} - \omega_{-{n}})(\omega_{+\sigma} - \omega_{+\ell}^*)(\omega_{+\sigma} -\omega_{- {m}}^*)} 
$   \\[1ex] \arrayrulecolor{white}\hline\\[-3ex]
$B_{\mu\nu}(x) = 2a_{\mu0}a_{\nu0}\mathrm{Im}[c_{\mu\nu}(x)],\quad
  \mat{B} = \int\!\! dx \,\mat{B}(x)
$
& 
 $\mat{Q}_{-\sigma} = \displaystyle\sum_{\ell{m}{n}}
\frac{\mat{P}_{+{n}}\mat{P}_{-\sigma} \mat{R}_-\mat{P}_{-\ell}^\dagger\mat{P}_{1{m}}^\dagger}
{(\omega_{-\sigma} - \omega_{+{n}}) (\omega_{-\sigma} - \omega_{+\ell}^*)(\omega_{-\sigma} - \omega_{+{m}}^*)} $
   \\[1ex]\hline\\[-3ex]
 $ \mat{M}_{\pm}\equiv
\pm \sqrt{\int\!\!dx \mat{A}(x)\gamma(x)}
+\frac{i}{2}\int\!\!dx \,\gamma(x)\,\mathbb{1}$
 &$\mat{S}^\sigma = 
\mathrm{Re}
\tfrac{2i}{a_0^2}
\Scale[0.9]{\left[\int \!\!dx\,\gamma(x)\mat{B}(x)\right]}
\left(\tfrac{\mat{Q}_{+\sigma}}{\omega_{+\sigma}^2} + \tfrac{\mat{Q}_{-\sigma}}{\omega_{-\sigma}^2}\right)
\Scale[0.9]{\left[\int \!\!dx\,\gamma(x)\mat{B}(x)\right]^T}
$ 
   \\[1.4ex]\hline\\[-3ex] 
 $ (\omega\mathbb{1} - \mat{M}_\pm)^{-1} = 
\displaystyle\sum_{\sigma}
\frac{\mat{P}_{\pm{\sigma}}}{\omega -\omega_{\pm\sigma}}
$&
$\mat{T}^{\sigma} = 
-\mathrm{Im}
\tfrac{2i}{a_0^2}
\Scale[0.9]{\left[\int \!\!dx\,\gamma(x)\mat{B}(x)\right]}
\left(\tfrac{\mat{Q}_{+\sigma}}{\omega_{+\sigma}^2} - \tfrac{\mat{Q}_{-\sigma}}{\omega_{-\sigma}^2}\right)
\Scale[0.9]{\left[\int \!\!dx\,\gamma(x)\mat{B}(x)\right]^T}
$\\[1.4ex]\hline\\[-3ex] 
 $ \omega _{\pm\sigma} =\pm\Omega_\sigma-i\Gamma_\sigma$&
$\mat{U}^\sigma = \mathrm{Re}\left(\tfrac{2i}{a_0^2}
\omega_{+\sigma}^2\mat{Q}_{+\sigma} + \omega_{-\sigma}^2\mat{Q}_{-\sigma}\right)
$\\[1.4ex]\hline\\[-3ex] 
  $\Gamma_{\mu\nu}\equiv
\tfrac{2[\mat{R}_0]_{\mu\mu}\delta_{\mu\nu}}{a_{\mu0}a_{\nu0}}+
\tfrac{2(\mat{B}\mat{A}^{-1}\mat{R}_0
{\mat{A}^\dagger}^{-1}\mat{B}^\dagger)_{\mu\nu}}
{a_{\mu0}a_{\nu0}}$ 
  &
$\mat{V}^\sigma = -\mathrm{Im}\tfrac{2i}{a_0^2}
\left(\omega_{+\sigma}^2\mat{Q}_{+\sigma} - \omega_{-\sigma}^2\mat{Q}_{-\sigma}\right)$
 \\[1.4ex]\hline\\[-3ex] 
 $\Gamma^{\mu\nu\sigma}_\mathrm{SB}\equiv \frac{\Gamma_{\mu\nu}}{2}+\Gamma_\sigma$ 
  &$\mat{X}^\sigma = 
\left[\int\!dx\, \gamma(x)^2\mat{B}(x)\right]
\tfrac{\mat{Q}_{+\sigma}}{\omega_{+\sigma}} + \tfrac{\mat{Q}_{-\sigma}}{\omega_{-\sigma}} 
$
\\[1.4ex]\hline\\[-3ex] 
 
&
$\mat{Y}^\sigma = 
2i\Scale[0.9]{\left[\displaystyle\int\!dx \,\gamma(x)^2\mat{B}(x)\right]}
\tfrac{\mat{Q}_{+\sigma}}{\omega_{+\sigma}} - \tfrac{\mat{Q}_{-\sigma}}{\omega_{-\sigma}} $
\\[1.4ex]\hline\\[-3ex] 
\end{tabular}}
\label{table:Factors1}
  \end{table}

For convenience, we summarize all the coefficients of  \eqref{Multimode-spectrum}  in Table 3. 
Similar to  \eqref{single-mode-result},
the first term represents the central peaks, which are Lorentzians  at the lasing-mode frequencies $\omega_\mu$, whose  widths $\Gamma_{\mu\mu}$  were derived in~\citeasnoun{pick2015ab}.
The second and third terms correspond to the $2M$ red and blue sidepeaks, associated with each lasing mode.
In contrast to the single-mode higher-order RO sidepeaks (mentioned above), which have exponentially decreasing intensities, the extra peaks  in the multimode case  have  comparable amplitudes  and should be measurable using standard experimental setups~\cite{vahala1983observation}.
The RO   frequencies and relaxation rates ($\Omega_\sigma$ and $\Gamma_\sigma$ respectively)  are obtained from  the real and imaginary parts of the  complex eigenvalues of the  matrix $\mat{M}$ (denoted  by $\omega_{\pm\sigma}$, with $\sigma = 1\hdots M$)\footnote{{ Since the matrix under the square root is positive definite, the square root is well defined. This point it justified in appendix B, following Eq. (B.16).}} . {  While $\Omega_\sigma$ determine the location of the RO peaks, $\Gamma_\sigma$ determine their linewidths, as can be seen from the definition of    $\Gamma^{\mu\nu\sigma}_\mathrm{SB}$ in Table 3. The projectors onto the eigenvectors of $\mat{M}_\pm$, which we label  in the table by $\mat{P}_{\pm\sigma}$, }
determine the multimode generalized $\alpha$ factors, which are expressed in terms of the matrices $\mat{S}^\sigma, \mat{T}^\sigma, \mat{U}^\sigma,\mat{V}^\sigma, \mat{X}^\sigma$ and $\mat{Y}^\sigma$.  Even though our derivation requires many pages of algebra, we compare the final result to numerical solution of  the nonlinear oscillator equations [\eqref{Multimode-TCMT}]  and the results match perfectly [\figref{numerical-TCMT}(c)].

\section{Discussion \label{discussion-sec}}

This paper  presented an \emph{ab-initio} formula for the   noise spectrum  of single- and multimode micro-structured complex-cavity lasers.   Our results are valid under very general conditions: ($i$) the laser having a stationary inversion and  reaching   a stable steady state,   ($ii$) operating far enough above the lasing threshold { (so that intensity fluctuations in each mode  are significantly smaller than the steady-state intensity),  ($iii$)  assuming that all the  lasing peaks and  sidebands are spectrally separated, and ($iv$) that spontaneous emission events are uncorrelated in time, which means that the noise autocorrelation function is treated as a constant within the spectral peaks (i.e., as white noise). 
}
As such, our theory is fairly general and accurately accounts for  inhomogeneity, cavity openness,  nonlinearity, and multimode effects in generic laser geometries. 
Since our formulas are expressed in terms of  the steady-state lasing modes and frequencies, their evaluation does not require substantial computation beyond solving the steady-state SALT equations (which can  be solved efficiently using  available algorithms~\cite{Tureci2006,Esterhazy2013}).

 We find a number of new effects, which arise from  the inhomogeneity of the lasing modes. 
 For example, we find    enhanced smearing and shifting of the RO sidepeaks in comparison to the traditional formulas (as demonstrated in~\figref{numerical-TCMT}), which follow from the spatial dependence of the effective  atomic-relaxation rate, $\gamma(\vec{x})$, above the lasing threshold. Additionally, we obtain three   generalized  $\alpha$ factors: the central-peak linewidth-enhancement factor, $\alpha_1$ (which was already presented  in \citeasnoun{pick2015ab}),  the  fractional power that goes into the sidepeaks,  $\alpha_2$, and the sideband-asymmetry factor,  $\alpha_3$.
We find that  $\alpha_1$  is always larger than the traditional factor, $\alpha_0$, while  $\alpha_2$ and $\alpha_3$ can be either larger or smaller than the traditional $\alpha_0$ (\figref{RO-properties}). The  generalized factors ($\alpha_{1,2,3}$) deviate significantly  from the  the traditional factor ($\alpha_0$)  in lasers with strong inhomogeniety, like random lasers~~\cite{He2013,Painter1999} or   lasers operating far above the threshold (where saturation effects become important). 

The theory in this paper can be applied to tackle additional    open questions in laser noise. For example, our current formulation treats only the effect of noise on the  modes above the lasing threshold, but understanding the  noise spectrum  near and slightly below the  threshold is   very important, e.g.,  in the study   of  light-emitting diodes (LEDs). Although there have been previous attempts to describe laser noise near the threshold~\cite{Hui1993}, the early theories use phenomenological rate equations for the lasing-mode amplitudes and  artificially interpolate  the sub-threshold and above-threshold regimes.  Along these lines, one could   interpolate \eqref{Osci-single} with the corresponding sub-threshold  equation and easily obtain an improvement over previous work, since the latter uses phenomenological rate equations while our generalized equations are obtained directly from Maxwell--Bloch. Another effect that  could potentially  be treated using our FDT-based approach, is the regime of strong amplified spontaneous emission (ASE), where noise from near-threshold modes can affect the steady-state lasing properties, i.e., by suppressing lasing due to taking up the gain. We anticipate that strong ASE could be treated by introducing an ensemble-averaged steady-state inversion, in which   noise from near-threshold modes would appear as an additional term in the gain saturation, where noise correlations are related to the steady-state properties of the medium by  the FDT. 
{  Additionally, one could straightforwardly generalize our approach to include  correlations between spontanteous emission events [relaxing assumption ($iv$) above], i.e., treat the random currents in \eqref{e-dot} as  colored noise. In the application of the residue theorem in the appendices, one would need to include residues that correspond to the poles of $R_{\mu\nu}(\omega)$, which are neglected in the current analysis.}
 These directions are further discussed in \citeasnoun{pick2017thesis}.

\section*{Funding}
This work was partially supported by the Army Research
Office through the Institute for Soldier Nanotechnologies
under Contract No. W911NF-13-D-0001.
AP  is partially supported  by an Aly Kaufman Fellowship at the Technion. 

\section*{Acknowledgments}
The authors would like to thank  A. Douglas  Stone for insightful discussion.

 
 \newpage
\appendix
\renewcommand{\theequation}{A.\arabic{equation}}
\setcounter{equation}{0}

\section*{Appendix A: Derivation of the single-mode noise spectrum  }

In this  appendix, we complete the  derivation of \eqref{single-mode-result} from the main text. After reviewing  some definitions from the main text in Sec.~A.1,  we calculate the second-order moments  of $u(t)$ and $\phi(t)$  in Sec.~A.2. Then, in Sec.~A.3, we use these results to obtain the power spectrum. 
 
\section*{A.1. Autocorrelations of the single-mode phase and intensity}

Recall that the Fourier transforms of $u(t)$,  $\phi(t)$, and $\xi(t)$  are [\eqref{linearSOLS}]:
\begin{subequations}
\begin{gather}
\tilde{\phi}(\omega) =
\frac{1}{i\omega + \int\!\!dx \frac{A(x)\gamma(x)}{\gamma(x)+ i\omega}}\cdot
\int\!\!dx
\tfrac{\gamma(x)B(x)}{\gamma(x) + i\omega}\cdot
\frac{\tilde{f}_R}{i\omega a_0}
+
\frac{\tilde{f}_I(\omega)}{i\omega a_0},
\label{eq:step3}
\\
\tilde{u}(\omega)
=\frac{1}{i\omega + 
\int\!\!dx \frac{A(x)\gamma(x)}{\gamma(x)+ i\omega}}\cdot
\frac{\tilde{f}_R}{a_0},
\label{eq:UAppend}
\\
\tilde{\xi}(x,\omega) =\frac{\gamma(x)}{\gamma(x) + i\omega}
\cdot\frac{1}{i\omega + \int\!\!dx \frac{A(x)\gamma(x)}{\gamma(x)+ i\omega}}\cdot
\frac{\tilde{f}_R}{a_0},
\label{eq:PhiUxiAppend}
\end{gather}
\label{eq:transforms-appendix}
\end{subequations}
\hspace{-4pt} where Fourier transforms are defined using the convention:
$\tilde{f}\equiv\frac{1}{\sqrt{2\pi}}\int_{-\infty}^\infty\!\!dt\,e^{-i\omega t}f(t)$~\cite{Arfken2006}.
Since intensity and phase  are  stationary random variables, the fluctuations at different frequencies are uncorrelated~\cite{Henry1986}
\begin{subequations}
\begin{align}
&\langle \tilde{\phi}(\omega)\tilde{\phi}^*(\omega')\rangle =
R_{\tilde{\phi} \tilde{\phi}}(\omega)\delta(\omega - \omega'),
\\
&\langle \tilde{u}(\omega)\tilde{u}^*(\omega')\rangle =
R_{\tilde{u} \tilde{u}}(\omega)\delta(\omega - \omega'),
\\
&\langle \tilde{\phi}(\omega)\tilde{u}^*(\omega')\rangle =
R_{\tilde{\phi} \tilde{u}}(\omega)\delta(\omega - \omega').
\end{align}
\end{subequations}
\hspace{-2pt}Given the autocorrelation of the Langevin noise $f$
\begin{gather}
\langle \tilde{f}(\omega)\tilde{f}^*(\omega) \rangle= R(\omega)\delta(\omega - \omega'),
\label{eq:step4}
\end{gather} 
[with $R$   given by the fluctuation dissipation theorem (in  Table 1)], 
and the explicit expressions for the Fourier transforms [\eqref{transforms-appendix}], we obtain
\begin{subequations}
\begin{gather}
R_{\tilde{\phi} \tilde{\phi}}(\omega) = 
\left(1+
\left|
\tfrac{\int\!\!dx
\tfrac{\gamma(x)B(x)}{\gamma(x) + i\omega}}{i\omega + \int\!\!dx \frac{A(x)\gamma(x)}{\gamma(x)+ i\omega}}\right|^2\,\right)\cdot \frac{ R(\omega)}{\omega^2 a_0^2},
\label{eq:R-phi-phi}
\\
R_{\tilde{u} \tilde{u}}(\omega) = 
\frac{1}{\left|i\omega + 
\int\!\!dx \frac{A(x)\gamma(x)}{\gamma(x)+ i\omega}\right|^2}\cdot
\frac{ R(\omega)}{a_0^2},
\label{eq:R-u-u}
\\
R_{\tilde{u} \tilde{\phi}}(\omega) = 
\frac{\int\!\!dx
\frac{\gamma(x)B(x)}{\gamma(x) + i\omega}}
{\left|i\omega + \int\!\!dx \frac{A(x)\gamma(x)}{\gamma(x)+ i\omega}\right|^2}\cdot
\frac{ R(\omega)}{i\omega a_0^2}.
\label{eq:R-phi-u}
\end{gather}
\end{subequations}
In the text we, show that the noise spectrum depends on the poles of  the autocorrelations in \eqref{transforms-appendix}. In order to find these poles, 
we introduce the   approximation  [\eqref{APPROX}]:
\begin{align}
\frac{1}{i\omega + \int\!\!dx \frac{A(x)\gamma(x)}{\gamma(x)+ i\omega}}   \approx
\frac{i\omega}{\int\!\!dx \left[i\omega (\gamma(x)+i\omega)+A(x)\gamma(x)\right]}=
\frac{-i\omega}{(\omega - \omega_+)(\omega - \omega_-)},
\label{eq:APPROXappend}
\end{align} 
which holds near the RO frequencies (i.e., when  $\omega\approx\Omega\gg\Gamma$). 
Using this approximation, one finds that the autocorrelations    have poles at  the complex RO frequencies
\begin{equation}
\omega_\pm \equiv  \pm\Omega+i\Gamma,
\end{equation}
where 
\begin{equation}
\Omega \equiv \Scale[0.9]{\sqrt{\int dx A(x)\gamma(x)}},
\quad\quad\quad
\Gamma \equiv \Scale[0.9]{\frac{1}{2}\int dx\gamma(x)}.
\end{equation}
Using \eqref{APPROXappend} and   $\pm i\omega+\gamma(x)\approx \pm i\omega$, we find that   the autocorrelations near the RO frequencies are:
\begin{subequations}
\begin{gather}
R_{\tilde{\phi} \tilde{\phi}}(\omega) \approx
\frac{ R(\omega)}{\omega^2 a_0^2}+
\left|
\frac{\int\!\!dx
\tfrac{\gamma(x)B(x)}{\gamma(x) + i\omega}}
{(\omega - \omega_+)(\omega - \omega_-)}\right|^2  \frac{ R(\omega)}{ a_0^2},
\label{eq:R-phi-phi-approx}
\\
R_{\tilde{u}\tilde{u}}(\omega)  \approx
\frac{\omega^2}{\left|(\omega - \omega_+)(\omega - \omega_-)\right|^2}\cdot
\frac{ R(\omega)}{a_0^2},
\label{eq:R-u-u-approx}
\\
R_{\tilde{u} \tilde{\phi}}(\omega) \approx
\frac{\omega^2\int\!\!dx
\tfrac{\gamma(x)B(x)}{\gamma(x) + i\omega}}
{\left|(\omega - \omega_+)(\omega - \omega_-)\right|^2}\cdot
\frac{ R(\omega)}{i\omega a_0^2}.
\label{eq:R-phi-u-approx}
\end{gather}
\end{subequations}

\section*{A.2. Second-order moments}

\section*{A.2.1.  The phase variance}

In the next section, we compute the phase variance by using its relation to the Fourier transform of the phase, $\tilde{\phi}(\omega) = \frac{1}{\sqrt{2\pi}}\int_{-\infty}^\infty dt e^{-i\omega t}\phi(t)$. 
In order to derive this relation  [\eqref{PhiVarFT} from the text], we write the phase difference in terms of the Fourier transform:
\begin{align}
\phi(t+t')\!-\!\phi(t')\!=\!\frac{1}{\sqrt{2\pi}}
\int{d}\omega \,\tilde{\phi}(\omega)\,e^{i\omega t'}(e^{i\omega t}\!-\!1).
\end{align}
Using this relation, we find that the phase variance equals 
\begin{gather}
\left<[\phi(t+t')-\phi(t')]^2 \right>=\frac{1}{2\pi}\iint\!\!{d}\omega \,{d}\omega' \left<\tilde{\phi}(\omega)\tilde{\phi}^*(\omega')\right>e^{i(\omega-\omega') t'}(e^{i\omega t}\!-\!1)(e^{-i\omega' t}\!-\!1)=\nonumber\\
\frac{1}{2\pi}\int\!\!{d}\omega \,R_{\tilde{\phi} \tilde{\phi}}(\omega)
(2\!-\!e^{i\omega t}\!-\!e^{-i\omega t}) = 
\mathrm{Re}\left[
\frac{1}{\pi}
\int_{-\infty}^\infty
{d}\omega \,
R_{\tilde{\phi} \tilde{\phi}}(\omega)
(1-e^{i\omega t})\right].
\label{eq:phi-correlations-appendix}
\end{gather} 
Substitution of the autocorrelation $R_{\tilde{\phi}\tilde{\phi}}$ [\eqref{R-phi-phi}] into \eqref{phi-correlations-appendix} yields
\begin{equation}
\left<[\phi(t+t')-\phi(t')]^2 \right>= 
\mathrm{Re}\left(
\frac{1}{\pi}
\int_{-\infty}^\infty
{d}\omega \,
\left[1+
\left|
\tfrac{\int\!\!dx
\tfrac{\gamma(x)B(x)}{\gamma(x) + i\omega}}{i\omega + \int\!\!dx \frac{A(x)\gamma(x)}{\gamma(x)+ i\omega}}\right|^2\,\right]\cdot 
\frac{ R(\omega) }{a_0^2}
\cdot
\frac{1-e^{i\omega t}}{\omega^2}\right)\equiv
\mathcal{J}_0 + \mathcal{J}_\pm,
\label{eq:define-Js}
\end{equation}
where we denote by   $\mathcal{J}_0$ and $\mathcal{J}_\pm$  the terms associated with the pole at $\omega = 0$ and at $\omega_\pm$ correspondingly. We compute the integrals by  performing analytic continuation into the complex plane (changing the integration variable from real $\omega$ to complex $z$) and  applying Cauchy's theorem~\cite{Arfken2006}.
{ 
The contribution of the pole at zero is
\begin{equation}
\mathcal{J}_0 =  \left(1+\left| \tfrac{ \int\! dx B(x) }{\int\! dx A(x)}\right|^2\right)\frac{ R(0)}{\pi a_0^2}\lim_{\beta\rightarrow0}\oint \tfrac{dz\,( 1-e^{iz t})}{(z+i\beta)(z-i\beta)},
\label{eq:prev-step}
\end{equation}
where we pulled outside  of the integral  the terms that d, and evaluated them at $z=0$. 
Next, we compute  the  integral by  moving the pole from $z=0$ away from the real axis~\cite{Arfken2006}: 
\begin{equation}
\int_{-\infty}^\infty\!\tfrac{d\omega \,( 1-e^{i\omega t})}{\omega^2}=\lim_{\beta\rightarrow0}
\oint \tfrac{dz\,( 1-e^{iz t})}{(z+i\beta)(z-i\beta)}=2\pi i 
\tfrac{ 1-e^{-\beta t}}{2i\beta} = \pi t.
\label{eq:last-step}
\end{equation}
Substituting \eqref{last-step} into \eqref{prev-step}, we obtain
\begin{gather}
\mathcal{J}_0   = \left[1+
\left| \tfrac{ \int\!dx\,B(x) }{\int\!dx\,A(x)}\right|^2\right]
\frac{\pi R(0)  t}{a_0^2}.
\end{gather}
The phase-drift coefficient is proportional to  $R(0)$, which is  determined by the gain at the lasing frequency, $\mathrm{Im}[\varepsilon(\vec{x},\omega_\mu)$.}
This term gives the central-peak linewidth with the $\alpha_1$-factor broadening. 

 Let us  denote the complex integrand by 
\begin{gather}
f(z)\equiv
\left[1+\left|
\tfrac{\int\!\!dx
\tfrac{\gamma(x)B(x)}{\gamma(x) + iz}}{iz + \int\!\!dx \frac{A(x)\gamma(x)}{\gamma(x)+ iz}}\right|^2\right]
\frac{ R(z)}{a_0^2}
\frac{ 1-e^{iz t}}{z^2},
\end{gather}
The RO terms are
\begin{align}
\mathcal{J}_\pm = 2\pi i\,\left[ \mathrm{Res}(f,\omega_+) + \mathrm{Res}(f,\omega_-)\right] .
\end{align}
In order to compute the residues  of the poles at $\omega_\pm$, we use the approximation for $R_{\tilde{\phi}\tilde{\phi}}(\omega)$ near the RO frequencies [\eqref{R-phi-phi-approx}] and   obtain
\begin{equation}
f(z)\approx
\frac{\left[\int\!dxB(x)\gamma(x)\right]^2}
{|(z-\omega_+)(z-\omega_-)|^2}
\frac{(1-e^{iz t})}{z^2}
\frac{R(\omega)}{a_0^2}
\end{equation}
where the residues at the complex RO frequencies are 
\begin{align}
\mbox{Res}(f,\omega_\pm)  = 
\frac{\Scale[0.9]{\left[\int\!dxB(x)\gamma(x)\right]^2} R(\omega_\pm)(1-e^{i\omega_\pm t})}
{a_0^2(\omega_\pm-\omega_\mp)(\omega_\pm-\omega_\pm^*)(\omega_\pm-\omega_\mp^*)\omega_\pm^2}\approx
 \Scale[1.2]{\frac{ \left[\int\!dx\,B(x)\gamma(x)\right]^2 }{\left[\int\!dx\,A(x)\gamma(x)\right]^2}}
\frac{R(\omega_\pm)(1-e^{i\omega_\pm t})}{4 \Gamma a_0^2}.
\end{align}
In the second equality,   we assumed that the sidebands are spectrally resolved from the main peak [i.e., that $\Omega\gg\Gamma$] and used the relation $\Omega^4 \approx  \left[\int\!dx\,A(x)\gamma(x)\right]^2$.
 { The amplitude of the RO sidepeaks is 
 proportional to  $R(\omega_\pm)$, which is  determined by the gain at the RO frequencies, $\mathrm{Im}[\varepsilon(\vec{x},\omega_\mu\pm\Omega)$.} Note that the gain and, hence, also $R(\omega)$ are symmetric functions around the lasing frequencies.  We introduce the shorthand notation: $R_0 \equiv R(0)$ and $R_\pm \equiv R(\omega_+) = R(\omega_-)$.  
Collecting the terms, we find:
\begin{align}
\boxed{\left<[\phi(t)-\phi(0)]^2 \right>= \frac{R_0}{a_0^2}\left(1+\alpha_1^2\right)t + 
\frac{R_\pm\alpha_2^2}{2a_0^2\Gamma}
\left(1-e^{-\Gamma t}\cos\Omega t\right)  -
\frac{3R_\pm\alpha_2^2}{2a_0^2\Omega} e^{-\Gamma t}\sin\Omega t}
\label{eq:box1}
\end{align}
where $\alpha_1 = \tfrac{ \int\!dx\,B(x) }{\int\!dx\,A(x)}$ and $\alpha_2 =  \tfrac{ \int\!dx\,B(x)\gamma(x) }{  
 \int\!dx\,A(x)\gamma(x)}$ are the first and second  generalized amplitude-phase couplings.

\section*{A.2.2.  Intensity autocorrelation}
Next, we apply similar tools to  compute  the autocorrelation of the intensity [\eqref{uVarResult}]. We begin by relating the  intensity autocorrelation to the  Fourier transform of the intensity:
\begin{align}
\left<[u(t+t') + u(t')]^2\right> = 
\mathrm{Re}\left[
\frac{1}{\pi}
\int_{-\infty}^\infty
{d}\omega 
R_{\tilde{u} \tilde{u}}(\omega)
(1+e^{i\omega t})
\right].
\label{eq:uVarAppend}
\end{align}
The Fourier-transformed intensity, $\tilde{u}$,  has poles only at the RO frequencies, $\omega_\pm$. We  approximate   $R_{\tilde{u}\tilde{u}}$  near the RO frequencies, and substitute \eqref{R-u-u-approx}  into \eqref{uVarAppend}. That yields   an improper integral that we calculate  using Cauchy's residue theorem:
\begin{gather}
\Scale[1.25]{\int_{-\infty}^\infty
{d}\omega 
\frac{\omega^2
(1+e^{i\omega t})}{|(\omega-\omega_+)(\omega-\omega_-)|^2}
=
\frac{2\pi i\omega_+^{2}(1+e^{i\omega_+t})}
{(\omega_+-\omega_-)(\omega_+-\omega_+^*)(\omega_+-\omega_-^*)}
+
\frac{2\pi i\omega_-^{2}(1+e^{i\omega_-t})}
{(\omega_--\omega_+)(\omega_--\omega_+^*)(\omega_--\omega_-^*)}}
\nonumber\\=
\Scale[1.25]{
\frac{\pi}{4\Omega\Gamma}
\left[
\frac{\omega_+^2(1+e^{i\omega_+t})}{\Omega+i\Gamma}
+
\frac{\omega_-^2(1+e^{i\omega_-t})}{\Omega-i\Gamma}
\right]}
\end{gather}
Substituting this result into \eqref{uVarAppend} and taking the limit of $\Omega\gg\Gamma$, we obtain \eqref{uVarResult} from the main text:
\begin{align}
\boxed{\left<[u(t) + u(0)]^2\right> =\frac{R_\pm}{2\Gamma a_0^2}
(1+\cos\Omega t e^{-\Gamma t})+
\frac{R_\pm}{2\Omega a_0^2}
\sin\Omega t e^{-\Gamma t}}
\label{eq:box2}
\end{align}

\section*{A.2.3. The cross term}
 
 Finally, let us compute  the time-averaged cross term by introducing   the Fourier transforms of $\tilde{u}$ and $\tilde{\phi}$. Using similar steps as in \eqref{phi-correlations-appendix}, we find:
\begin{align}
\left<[\phi(t+t')-\phi(t')][u(t+t') + u(t')]\right> = 
\frac{1}{2\pi}
\int_{-\infty}^\infty
{d}\omega \,
\left(e^{i\omega t} - e^{-i\omega t}\right)R_{\tilde{\phi}\tilde{u}}.  
\label{eq:CrossTermAppend}
\end{align}
We substitute the autocorrelation $R_{\tilde{\phi}\tilde{u}}$ [\eqref{R-phi-u}] into \eqref{CrossTermAppend}. The resulting expression has  poles at $\omega = 0$ and at $\omega_\pm$, and we denote their contributions by $\mathcal{I}_0$ and $\mathcal{I}_\pm$ respectively:
\begin{gather}
\left<[\phi(t)-\phi(0)][u(t) + u(0)]\right> = 
\int_{-\infty}^\infty
{d}z 
\left(
\tfrac{R}{2\pi i a_0^2}\cdot
\tfrac{\int\!\!dx\,\tfrac{\gamma(x)B(x)}{\gamma(x) + i\omega}}{\left|i\omega + \int\!\!dx \frac{A(x)\gamma(x)}{\gamma(x)+ i\omega}\right|^2}
\cdot\left[\tfrac{e^{i\omega t} - e^{-i\omega t}}{\omega}\right]
\right) \equiv\mathcal{I}_0+\mathcal{I}_\pm
\label{eq:define-Is}
\end{gather}
We use standard results from complex analysis~\cite{Arfken2006} to compute the residue of the pole at $\omega = 0$ and find
\begin{gather}
\mathcal{I}_0 =\frac{R_0}{a_0^2}\cdot\frac{B}{A^2}.
\end{gather}
The contribution of the poles at $\omega_\pm$ can be found  by approximating  $R_{\tilde{\phi}\tilde{u}}$  near the RO frequencies [\eqref{R-phi-u-approx}]:
\begin{equation}
R_{\tilde{\phi}\tilde{u}} \approx
\frac{R_\pm}{a_0^2}\,
\int\!dx\, \frac{B(x)\gamma(x)}{\gamma(x)^2+\omega^2}\cdot
\frac{\omega^2\left(\frac{\gamma(x)}{i\omega} - 1\right)	}{\, 
|(\omega-\omega_+)(\omega - \omega_-)|^2}.
\label{eq:CrossTermStep}
\end{equation}
When substituting this result into \eqref{CrossTermAppend}, it becomes  apparent that only the odd part of $R_{\tilde{\phi}\tilde{u}}$ contributes to the integral since $(e^{i\omega t} - e^{-i\omega t})$ is an odd function in $\omega$ . Therefore, we  replace the term  $\left[\frac{\gamma(x)}{i\omega} - 1\right]$ in the numerator of the integrand by $\frac{\gamma(x)}{i\omega}$ and obtain
\begin{gather}
\mathcal{I}_\pm  = 
\frac{R_\pm}{\pi a_0^2i}\,
\int\!dx\, {B(x)}
\int_{-\infty}^\infty
{d}\omega 
\frac{e^{i\omega t}}{\omega}
\frac{\gamma(x)^2}{\gamma(x)^2+\omega^2}
\frac{\omega^2}{ |(\omega-\omega_+)(\omega-\omega_-)|^2}\nonumber\\
=\Scale[1.2]{R_\pm \frac{\int\!dx\,B(x)\gamma(x)^2}{\pi i a_0^2}\cdot
\left[
\frac{2\pi i  \,e^{i\omega_+ t}}
{\omega_+(\omega_+ - \omega_-)(\omega_+-\omega_+^*)(\omega_+-\omega_-^*)} +
\frac{2\pi i  \,e^{i\omega_- t}}
{\omega_-(\omega_- - \omega_+)(\omega_--\omega_-^*)(\omega_--\omega_+^*)}
\right]=}
\nonumber\\
\Scale[1.2]{R_\pm\frac{\int\!dx\,B(x)\gamma(x)^2}{4i\Gamma\Omega a_0^2}
\left[
\frac{e^{i\omega_+t}}{(\Omega+i\Gamma)^2}-\frac{e^{i\omega_-t}}{(\Omega-i\Gamma)^2}
\right]},
\label{eq:Ipm}
\end{gather}
where in going from the first to second line, we used the residue theorem, and in going from the second to third line, we substituted $\omega_\pm = \pm\Omega-i\Gamma$.
Collecting these results, we obtain
\begin{align}
\boxed{\left<[\phi(t)-\phi(0)][u(t) + u(0)]\right>= \tfrac{R_0\,\alpha}{a_0^2\,A}
+\tfrac{R_\pm\,\alpha_3}{a_0^2\Omega}\left(
-\tfrac{2\Gamma}{\Omega}
\cos\Omega t\,e^{-\Gamma t}+\sin\Omega t\,e^{-\Gamma t}\right)}
\label{eq:box3}
\end{align}
where the definition of  $\alpha_3$ is given in Table 2.

\section*{A.3. The power spectrum}

In this section, we  derive  a simplified formula for the autocorrelation,  $\langle a(t)a^*(0)\rangle$. Then, we compute  its  Fourier transform  and obtain the  single-mode noise spectrum formula [\eqref{single-mode-result} from the main text]. In order to simplify the notation, we introduce the parameters  $w_1, w_2,\hdots,w_8$ and rewrite   the second-order moments from Sec.~A.2 in the form:
\begin{subequations}
 \begin{gather}
\left<[\phi(t)-\phi(0)]^2\right>  = w_1 t+ w_2(1-e^{-\Gamma t}\cos\Omega t) + w_3e^{-\Gamma t}\sin\Omega t
\label{eq:PhaseCorrFinal}
\\ 
\left<[u(t)+u(0)]^2\right>  = w_4(1+e^{-\Gamma t}\cos\Omega t)  + w_5 e^{-\Gamma t}\sin\Omega t
\label{eq:IntensityCorrFinal}
\\ 
\left<[u(t)+u(0)][\phi(t)-\phi(0)]\right>  = 
\,w_6+ w_7e^{-\Gamma t}\cos\Omega t + aw8e^{-\Gamma t}\sin\Omega t
\label{eq:PhaseIntensityCorr}
\end{gather}
\label{eq:autocorrelatiopns-formal}
\end{subequations}
\hspace{-3pt}We substitute these expressions into the autocorrelation of $a$ [\eqref{autocorrelation} from the main text,  restated here for convenience]:
\begin{align}
\frac{\left<a(t+t')a^*(t')\right>}
{\left<|a(t')|^2\right>} = 
e^{
-\frac{1}{2}
\left\{
\left<[\phi(t+t')-\phi(t')]^2\right>
-\left<[u(t+t')+u(t')]^2\right>+4\left<[u(t')]^2\right>
 \right\}-i
\left<[u(t+t')+u(t')][\phi(t+t')-\phi(t')]\right>
}.
\label{eq:autoAppendix}
\end{align}
Next, we introduce  an approximation that makes the power spectrum analytically solvable: When the RO terms in  \eqref{autocorrelatiopns-formal} are small (i.e., when $w_2,\hdots,w_8\ll1$), one can expand the corresponding exponential factors  in \eqref{autoAppendix}  in a Taylor series (e.g.,    $e^{w_2}\approx 1+w_2$ etcetera).  
In this regime, we find
\begin{align}
&\tfrac{\left<a(t+t')a^*(t')\right>}
{\left<|a(t')|^2\right>}  \approx\nonumber\\
&\begin{cases}
e^{-\tfrac{w_1|t|}{2}}\left(1-\tfrac{w_2+w_4+2iw_6}{2}\right) +
e^{-\Gamma_\mathrm{eff}|t|}
\left[\cos\Omega |t| 
\left(\tfrac{w_2+w_4-2iw_7}{2}\right) +\sin\Omega |t| 
\left(\tfrac{w_5-w_3-2iw_8}{2}\right)\right]\\
 \hspace{4in} \text{if $t>0$}\\
e^{-\tfrac{w_1|t|}{2}}\left(1-\tfrac{w_2+w_4-2iw_6}{2}\right)  +
\left[ 
e^{-\Gamma_\mathrm{eff}|t|}\cos\Omega |t| 
\left(\tfrac{w_2+w_4+2iw_7}{2}\right)
+\sin\Omega |t| 
\left(\tfrac{w_5-w_3+2iw_8}{2}\right)\right]
 \\
 \hspace{3.9in}
  \text{otherwise},
  \end{cases}
  \label{eq:TransformMe}
 \end{align} 
where $\Gamma_\mathrm{eff}\equiv\tfrac{w_1}{2}+\Gamma$.
The spectrum is then found by taking  the Fourier transform of \eqref{TransformMe}. 
After some algebra, we obtain
\begin{align}
&S(\omega) = \tfrac{w_1}{\omega^2+(w_1/2)^2}
\left(1-\tfrac{w_2+w_4+2w_6}{2}\right) +\nonumber\\
& 
\tfrac{\Gamma_\mathrm{eff}}{(\omega+\Omega)^2+\Gamma_\mathrm{eff}^2}
\left[
\left(\tfrac{w_2+w_4+2w_8}{2}\right)
+
\tfrac{\Omega+\omega}{\Gamma_\mathrm{eff}}
\left(\tfrac{w_5-w_3+2w_7}{2}\right)
\right]+\tfrac{\Gamma_\mathrm{eff}}{(\omega-\Omega)^
2+\Gamma_\mathrm{eff}^2}
\left[
\left(\tfrac{w_2+w_4-2w_8}{2}\right)
-
\tfrac{\Omega-\omega}{\Gamma_\mathrm{eff}}
\left(\tfrac{w_5-w_3-2w_7}{2}\right)
\right].
\end{align} 
By comparing \eqref{autocorrelatiopns-formal} with the boxed equations from the previous section [\eqref{box1}, \eqref{box2}, and \eqref{box3}], we find the coefficients:
\begin{gather}
w_1 = \tfrac{R_0\left(1+\alpha_1^2\right)}{a_0^2}, 
w_2 = \tfrac{R_\pm\alpha_2^2}{2a_0^2\Gamma},  
w_3 = -\tfrac{3R_\pm\alpha_2^2}{2a_0^2\Omega},
w_4  = \tfrac{R_\pm}{2\Gamma a_0^2},\nonumber\\
 w_5 =\tfrac{R_\pm}{2\Omega a_0^2},
w_6 = \tfrac{R_0\,\alpha_1}{a_0^2A},
w_7 = \tfrac{2\Gamma R_\pm\,\alpha_3}{a_0^2\Omega^2}, 
 w_8 = \tfrac{R_\pm\,\alpha_3}{a_0^2\Omega} 
\label{eq:coeffs}
\end{gather}
Note that the  RO terms in  \eqref{autocorrelatiopns-formal}  are indeed small when $R(1+\alpha_1^2)\ll\Gamma$ and our approximation in \eqref{TransformMe} is legitimate. 
 formulaThat completes the derivation of the single-mode noise-spectrum formula.




\renewcommand{\theequation}{B.\arabic{equation}}
\setcounter{equation}{0}

\section*{Appendix B: Derivation of the multimode formula\label{appendixMultimode}}

\section*{B.1. Multimode oscillator equation}

In this appendix, we compute the sideband spectrum for a multimode laser. We showed in~\citeasnoun{pick2015ab} that the mode amplitudes  obey coupled   nonlinear oscillator equations:
\begin{align}
\Scale[0.95]{\dot{a}_\mu = \sum_{\nu k } C_{\mu\nu}^k  
\left[\gamma_k \int^t\!\!
dt'e^{-\gamma_k(t-t')}\left(a_{\nu0}^2 - |a_\nu(t')|^2\right)\right]a_\mu + f_\mu}
\label{eq:multi-osci}
\end{align}
Here, $\mu,\nu = 1, \hdots, M$, where  $M$ is the number of  lasing modes and 
$k = 1, \hdots, N$, where $N$  is the number of grid points   (when discretizing  space, e.g., by employing a finite-difference approach or a Riemann sum). At the end of the derivation, we take the limit of $N\rightarrow\infty$, obtaining results which  are independent of the discretization (similar to the approach of~\citeasnoun{pick2015ab}).  Similar to the analysis of the single-mode case, we separate the intensity and phase deviations of the modal amplitudes: 
\begin{equation}
a_\mu = a_{\mu0}e^{-u_\mu + i\phi_\mu}.
\label{eq:multi-ansatz}
\end{equation}
The multimode autocorrelation is
\begin{align}
&\left<a_\mu(t+t')a_\nu^*(t')\right> = \nonumber\\
&\Scale[0.9]{
\exp\left[{
-\frac{1}{2}
\left\{
\underbrace{\left<[\phi_\mu(t+t')-\phi_\mu(t')][\phi_\nu(t+t')-\phi_\nu(t')]\right>}_{\text{phase variance}}
-\underbrace{\left<[u_\mu(t+t')+u_\mu(t')][u_\nu(t+t')+u_\nu(t')]\right>}_{\text{intensity autocorrelation}}
 \right\}}\right] \times }
 \nonumber\\
&\Scale[0.9]{\exp\left[{-i
\underbrace{\left<[u_\mu(t+t')+u_\mu(t')][\phi_\nu(t+t')-\phi_\nu(t')]\right>}_{\text{cross term}}
}\right]\times\underbrace{\exp[{i\omega_\mu t}]}_{\text{lasing frequency}}.}
\label{eq:multi-auto}
\end{align}
In order to compute the second-order moments of $u_\mu$ and $\phi_\mu$, we substitute \eqref{multi-ansatz} into \eqref{multi-osci} and  linearize the equations around the  steady state (i.e., assuming small intensity fluctuations, $u_\mu\ll a_{\mu0}$). We  obtain
\begin{subequations}
\begin{gather}
\dot{u}_\mu = -\sum_{\nu k}A_{\mu\nu}^k\xi_\nu^k + f_\mu^R
\label{eq:multi-u}
\\
\dot{\phi}_\mu = \sum_{\nu k}B_{\mu\nu}^k\xi_\nu^k + f_\mu^I
\label{eq:multi-phi}
\\
\dot{\xi}_\mu^k = -\gamma_k{\xi}_\mu^k + \gamma_ku_\mu,
\label{eq:multi-xi}
\end{gather}
\label{eq:multi-linearized}
\end{subequations}
\hspace{-4pt}where $\xi_\mu^k = \gamma_k\int^t dt' e^{-\gamma_k(t-t')}u_\mu(t')$ is the time-delayed intensity fluctuation while 
 $A_{\mu\nu}^k\equiv2a_{\nu0}^2\mathrm{Re}[C_{\mu\nu}^k]$ and $B_{\mu\nu}^k\equiv2a_{\nu0}^2\mathrm{Re}[C_{\mu\nu}^k]$ are the real and imaginary parts of the nonlinear-coupling matrix $C_{\mu\nu}^k$. 
Similar to the single-mode case, we proceed by taking  the Fourier transforms of \eqref{multi-linearized}. 
First, we solve the set of equations for $\tilde{u}$ and $\tilde{\xi}_k$ and then use the results to compute $\tilde{\phi}$.  We begin by rewriting the equations for $\tilde{u}_\mu$ and  $\tilde{\xi}_\mu^k$ in matrix form,
\begin{equation}
\tilde{\vec{x}} = [i\omega\mathbb{1} + \bold{K}]^{-1}\tilde{\vec{f}},
\label{eq:linear}
\end{equation}
where
\begin{equation}
\Scale[0.9]{
\bold{K}=
\left( \begin{array}{cccc}
0    & \mat{A}_1 & \hdots & \mat{A}_N \\
-\gamma_1\mathbb{1} & \gamma_1\mathbb{1} & & \\
\vdots & & \ddots& \\
-\gamma_N\mathbb{1} & & &\gamma_N\mathbb{1}\end{array} \right)
\hspace{0.2in}
\tilde{\vec{x}}=
\left( \begin{array}{c}
\tilde{\vec{u} }\\
\tilde{\vecg{\xi}}^1\\
\vdots \\
\tilde{\vecg{\xi}}^N   \end{array} \right)
\hspace{0.2in}
\tilde{\vec{f}}=
\left( \begin{array}{c}
\tilde{\vec{f}}_R \\
0  \\
\vdots\\
0 \end{array} \right).}
\end{equation}
$\tilde{\vec{u}}$, $\tilde{\vec{f}}_R$, and $\tilde{\vecg{\xi}}^k$ are vectors  whose entries are $\tilde{u}_\mu$, $\mathrm{Re}\,[\tilde{f}_\mu]$, and $\tilde{\xi}^k_\mu$ respectively. The symbol  $\mathbb{1}$ denotes the $M\times M$ identity matrix and $\mat{A}_k$ is the $M\times M$ matrix $\mat{A}_k = 2a_0^2\mat{C}_k$.
In order to solve \eqref{linear} and find $\tilde{\vec{u}}$ and $\tilde{\vecg{\xi}}_k$, we need to invert the matrix   $(i\omega\mathbb{1} + \bold{K}) $ which we can write formally as  
\begin{equation}
i\omega\mathbb{1} + \bold{K} =
\left( \begin{array}{cc}
\mat{X}   & \mat{Y} \\
\mat{Z}  & \mat{W}\end{array} \right).
\end{equation}
Here 
\begin{gather}
\mat{X} = \mathbb{1}\omega
\hspace{1in}\mat{Y}=
\left( \begin{array}{ccc}
\mat{A}_1   & \hdots & \mat{A}_N\end{array} \right)
\nonumber\\
\Scale[0.9]{
\mat{Z}=
\left( \begin{array}{c}
-\gamma_1\mathbb{1}   \\
\vdots  \\
-\gamma_N\mathbb{1}  \end{array} \right)
\hspace{0.5in}\mat{W}=
\left( \begin{array}{ccc}
\gamma_1\mathbb{1} + i\omega \mathbb{1} & & \\
 &\ddots&  \\
 & &\gamma_N\mathbb{1}  + i\omega \mathbb{1}\end{array} \right)}.
 \end{gather}
Using Schur's complement~\cite{Bultheel1997}, the matrix inverse is
\begin{align}
\Scale[0.9]{[i\omega\mathbb{1} + \bold{K}]^{-1}=}
\Scale[0.8]{
\left( \begin{array}{cc}
(\mat{X}-\mat{Y}\mat{W}^{-1}\mat{Z})^{-1}   & -(\mat{X}-\mat{Y}\mat{W}^{-1}\mat{Z})^{-1}\mat{Y}\mat{W}^{-1} \\
-\mat{W}^{-1}\mat{Z}(\mat{X}-\mat{Y}\mat{W}^{-1}\mat{Z})^{-1}   & \mat{W}^{-1} + \mat{W}^{-1}\mat{Z}(\mat{X}-\mat{Y}\mat{W}^{-1}\mat{Z})^{-1} \mat{Y}\mat{W}^{-1}\end{array} \right)}
\end{align}
Therefore, we obtain
\begin{gather}
\tilde{\vec{u}} = (\mat{X}-\mat{Y}\mat{W}^{-1}\mat{Z})^{-1}\frac{\tilde{\vec{f}}_R}{a_0}
\nonumber\\
\tilde{\vecg{\xi}}^k = -\left[\mat{W}^{-1}\mat{Z}(\mat{X}-\mat{Y}\mat{W}^{-1}\mat{Z})^{-1}\right]_{k}\frac{\tilde{\vec{f}}^R}{a_0} 
\end{gather}
where $[\mat{O}]_k$  denotes the $k$th block of the matrix $\mat{O}$. We obtain explicit expressions:
\begin{subequations}
\begin{gather}
\tilde{\vecg{\phi}} =\frac{ \sum_k \mat{B}_k\,\vecg{\xi}^k}{i\omega}+\frac{\tilde{\vec{f}}_I}{i\omega}
\label{eq:multi-phi}
\\
\tilde{\vec{u}} = 
\left({i\omega\mathbb{1} + \sum_k  \tfrac{\gamma_k \mat{A}_k }{\gamma_k+ i\omega}
}\right)^{-1}
\frac{\tilde{\vec{f}}_R}{a_0}
\label{eq:multi-u}\\
\tilde{\vecg{\xi}}^k =  
\tfrac{\gamma_k}{\gamma_k+ i\omega}
\cdot
\left({i\omega\mathbb{1} + \sum_k  
\tfrac{\gamma_k \mat{A}_k }{\gamma_k+ i\omega}
}\right)^{-1}
\frac{\tilde{\vec{f}}^R}{a_0}
\label{eq:multi-xi}
\end{gather}
\label{eq:multi-fourier-sols}
\end{subequations}

\section*{B.2 Autocorrelations of the multimode phase and intensity}

The multimode matrix autocorrelations are defined as
\begin{subequations}
\begin{gather}
\left<\tilde{\vecg{\phi}}(\omega)\tilde{\vecg{\phi}}^\dagger(\omega')\right>  =
\mat{R}_{\tilde{\vecg{\phi}}\tilde{\vecg{\phi}}}(\omega)\delta(\omega - \omega')\\
\left<\tilde{\vec{u}}(\omega)\tilde{\vec{u}}^\dagger(\omega')\right>  =
\mat{R}_{\tilde{\vec{u}}\tilde{\vec{u}}}(\omega)\delta(\omega - \omega')\\
\left<\tilde{\vecg{\phi}}(\omega)\tilde{\vec{u}}^\dagger(\omega')\right>  =
\mat{R}_{\tilde{\vecg{\phi}}\tilde{\vec{u}}}(\omega)\delta(\omega - \omega'),
\end{gather}
\end{subequations}
where
\begin{subequations}
\begin{gather}
\Scale[0.9]{
\mat{R}_{\tilde{\vecg{\phi}}\tilde{\vecg{\phi}}} (\omega)= 
\sum_{k\ell} \frac{\gamma_k\mat{B}_k}{\gamma_k + i\omega}\cdot
\left({i\omega\mathbb{1} + \sum_k  \tfrac{\gamma_k \mat{A}_k }{\gamma_k+ i\omega}
}\right)^{-1}\frac{\mat{R}(\omega)}{\omega^2 a_0^2}
\left({-i\omega\mathbb{1} + \sum_k  \tfrac{\gamma_k \mat{A}_k^\dagger }{\gamma_k- i\omega}
}\right)^{-1}\cdot
 \frac{\gamma_\ell\mat{B}_\ell^\dagger}{\gamma_\ell - i\omega} + 
 \frac{\mat{R}(\omega)}{\omega^2 a_0^2}}
 \\
\Scale[0.9]{\mat{R}_{\tilde{\vec{u}}\tilde{\vec{u}}}(\omega)  = 
\left({i\omega\mathbb{1} + \sum_k  \tfrac{\gamma_k \mat{A}_k }{\gamma_k+ i\omega}
}\right)^{-1}\frac{\mat{R}(\omega)}{a_0^2}
\left({-i\omega\mathbb{1} + \sum_k  \tfrac{\gamma_k \mat{A}_k^\dagger }{\gamma_k- i\omega}
}\right)^{-1}}\\
\Scale[0.9]{\mat{R}_{\tilde{\vecg{\phi}}\tilde{\vec{u}}}(\omega) = 
\sum_{k} \tfrac{\gamma_k\mat{B}_k}{\gamma_k + i\omega}\cdot
\left({i\omega\mathbb{1} + \sum_k  \tfrac{\gamma_k \mat{A}_k }{\gamma_k+ i\omega}
}\right)^{-1}\frac{\mat{R}(\omega)}{i\omega a_0^2}
\left({-i\omega\mathbb{1} + \sum_k  \tfrac{\gamma_k \mat{A}_k^\dagger }{\gamma_k- i\omega}
}\right)^{-1}}.
 \label{eq:multi-R-phiu}
\end{gather}
\end{subequations}

In the next section, we compute the second-order moments for $\phi_\mu$ and $u_\mu$. As in the single-mode case, the result will depend on the poles of the Fourier transforms. We find that the Fourier transforms have  poles at $\omega = 0$ and $2M$ additional poles for each lasing mode, which give rise to  $2M$ RO sidepeaks \emph{around each lasing frequency}. In order to see this, we  rewrite  the matrix $\left({i\omega\mathbb{1} + \sum_k 
\tfrac{\gamma_k \mat{A}_k }{\gamma_k+ i\omega}}
\right)$ in a way that easily shows the frequencies $\omega$ for which the matrix is null.
Similar to \eqref{APPROX}, we use the approximation near the RO frequencies (the validity regime will be checked at the end)
\begin{equation}
\Scale[1]{{i\omega\mathbb{1} + \sum_k \tfrac{\gamma_k \mat{A}_k }{\gamma_k+ i\omega}}
=\displaystyle\sum_k 
\left[i\omega(\gamma_k +i\omega)\mathbb{1}+\mat{A}_k \gamma_k \right]
\frac{1}{\gamma_k  +i\omega }\approx
\tfrac{1}{i\omega}\displaystyle\sum_k 
\left[i\omega(\gamma_k +i\omega)\mathbb{1}+\mat{A}_k \gamma_k \right]}.
\label{eq:multi-approx}
\end{equation}
The term in square brackets is a second-degree matrix polynomial in $\omega$, which can be rewritten as 
\begin{equation}
i\omega\displaystyle_k \gamma_k\mathbb{1} -\omega^2\mathbb{1}+\displaystyle\sum_k \mat{A}_k \gamma_k  = -(\omega\mathbb{1}-\mat{M}_+)(\omega\mathbb{1}-\mat{M}_-),
\label{eq:matrix-quadratic}
\end{equation}
where we introduced the definition
\begin{equation}
\mat{M}_{\pm} = \pm\sqrt{ \sum_k \gamma_k \mat{A}_k   - \left(\tfrac{1}{2}\sum_k  \gamma_k\mathbb{1} \right)^2} + \tfrac{i}{2}\sum_k  \gamma_k\mathbb{1} .
\end{equation}
The square root of a diagonalizable matrix $\mat{O} = \mat{V}\mat{D}\mat{V}^{-1}$ is
 $\sqrt{\mat{O} } = \mat{V}\sqrt{\mat{D}}\mat{V}^{-1}$. Note that  the matrix  $\sum_k \mat{A}_k \gamma_k  - \left(\tfrac{1}{2}\sum_k  \gamma_k\mathbb{1} \right)^2$ is positive definite because (1) The matrices $\mat{A}_k$ are  positive definite, as this is a stability criterion for \eqref{multi-osci}, and (2) $\norm{\mat{A}_k}>\gamma_k$ (where  $\norm{,}$ is a matrix norm), as this is a stability criterion for SALT (i.e., SALT assumes a steady-state inversion, and that requires small atomic relaxation rates).
Substituting \eqref{multi-approx} and \eqref{matrix-quadratic} in \eqref{multi-fourier-sols}, we obtain approximate expressions for the Fourier transforms:
\begin{subequations}
\begin{gather}
\tilde{\vecg{\phi}}(\omega) \approx\sum_k
\tfrac{\gamma_k\mat{B}_k}{\gamma_k + i\omega}
\left[(\omega\mathbb{1}-\mat{M}_+)(\omega\mathbb{1}-\mat{M}_-)\right]^{-1}\frac{\tilde{\vec{f}}^R(\omega)}{a_0}
+\frac{\tilde{\vec{f}}_I(\omega)}{i\omega}
\\
\tilde{\vec{u}}(\omega) \approx i\omega
\left[(\omega\mathbb{1}-\mat{M}_+)(\omega\mathbb{1}-\mat{M}_-)\right]^{-1}
\frac{\tilde{\vec{f}}_R(\omega)}{a_0}
\\
\tilde{\vecg{\xi}}^k(\omega) \approx  
\tfrac{\gamma_k}{\gamma_k+ i\omega}
\cdot
 i\omega
\left[(\omega\mathbb{1}-\mat{M}_+)(\omega\mathbb{1}-\mat{M}_-)\right]^{-1}\frac{\tilde{\vec{f}}^R(\omega)}{a_0}
\end{gather}
\end{subequations}
In order to find the location of the poles in  the integrand of \eqref{multi-phi-correlation}, we introduce the  eigenvalue decomposition of the resolvent operator, $\mat{M}_+$ and $\mat{M}_-$: 
\begin{align}
(\omega \mathbb{1}- \mat{M}_\pm)^{-1} = 
\sum_{\sigma}
\frac{\mat{P}_{\pm{\sigma}}}{\omega - \omega _{\pm\sigma}}
\label{eq:multi-decomposition}
\end{align}
where $i\omega_{\pm\sigma}$ are the eigenvalues of $\mat{M}_\pm$ and $\mat{P}_{\pm\sigma}$ are projection operators onto the corresponding eigenspaces.  The real and imaginary parts of $\omega_{\pm\sigma}$ determine the frequencies and widths of the RO side peaks. 
Using this approximation [\eqref{matrix-quadratic}], we can approximate the multimode Fourier transforms near the RO frequencies:
\begin{subequations}
\begin{gather}
\mat{R}_{\tilde{\vecg{\phi}}\tilde{\vecg{\phi}}}(\omega) \approx
 \frac{1}{a_0^2}\cdot
\displaystyle\sum_{\substack{k\ell \\ \mu\nu\sigma\tau}}
\frac{\gamma_k\mat{B}_k}{\gamma_k + i\omega}
\frac{\mat{P}_{-\mu}\mat{P}_{+\nu}\mat{R}(\omega)\mat{P}^\dagger_{+\sigma}\mat{P}_{-\tau}}
{(\omega-\omega_{-\mu})(\omega - \omega_{+\nu})(\omega - \omega_{+\sigma}^*)(\omega-\omega_{-\tau}^*)}
 \frac{\gamma_\ell\mat{B}_\ell^\dagger}{\gamma_\ell - i\omega} + 
 \frac{\mat{R}(\omega)}{\omega^2 a_0^2}
 \label{eq:multi-R-phiphi-approx}
  \\
\mat{R}_{\tilde{\vec{u}}\tilde{\vec{u}}}(\omega)  \approx 
\frac{\omega^2}{a_0^2}\cdot
\displaystyle\sum_{ \mu\nu\sigma\tau}
\frac{\mat{P}_{-\mu}\mat{P}_{+\nu}\mat{R}(\omega)\mat{P}^\dagger_{+\sigma}\mat{P}_{-\tau}}
{(\omega-\omega_{-\mu})(\omega - \omega_{+\nu})(\omega - \omega_{+\sigma}^*)(\omega-\omega_{-\tau}^*)}
 \label{eq:multi-R-uu-approx}\\
\mat{R}_{\tilde{\vecg{\phi}}\tilde{\vec{u}}}(\omega) \approx 
\sum_{k\mu\nu\sigma\tau}
\frac{\gamma_k\mat{B}_k}{\gamma_k + i\omega}
\frac{\mat{P}_{-\mu}\mat{P}_{+\nu}\mat{R}(\omega)\mat{P}^\dagger_{+\sigma}\mat{P}_{-\tau}}
{(\omega-\omega_{-\mu})(\omega - \omega_{+\nu})(\omega - \omega_{+\sigma}^*)(\omega-\omega_{-\tau}^*)}\cdot 
\frac{\omega}{i a_0^2} 
 \label{eq:multi-R-phiu-approx}
\end{gather}
\end{subequations}

\section*{B.3. The multimode second-order moments}

\section*{B.3.1 The phase variance}
Similar to the derivation from Sec.~A.2.1, we relate the multimode phase variance to the autocorrelation of the phases
\begin{gather}
\left<\Scale[0.95]{
[\vecg{\phi}(t+t') - \vecg{\phi}(t')] [\vecg{\phi}^T\!(t+t') - \vecg{\phi}^T\!(t')]}\right> =
\Scale[0.98]{
\mathrm{Re}\left[
\tfrac{1}{\pi}\int_{-\infty}^\infty
d\omega 
\mat{R}_{\tilde{\vecg{\phi}}\tilde{\vecg{\phi}}}(\omega)
[1-e^{i\omega t}]\right] \equiv\mat{J}_0 + \mat{J}_\pm.}
\label{eq:multi-phi-correlation}
\end{gather}
where in the last equality we separate the contributions of the poles at $\omega = 0$ and the poles associated with RO dynamics. From Sec.~B.2, the phase autocorrelation is
\begin{gather}
\mat{R}_{\tilde{\vecg{\phi}}\tilde{\vecg{\phi}}}(\omega) =
\displaystyle\sum_{k\ell}
\tfrac{\mat{B}_k\gamma_k}{\gamma_k+i\omega}
\left(i\omega\mathbb{1} + \sum_k \tfrac{\gamma_k \mat{A}_k }{\gamma_k+ i\omega}\right)^{-1}
\tfrac{\mat{R}(\omega)}{\omega^2 a_0^2}
\left(-i\omega\mathbb{1} + \sum_k \tfrac{\gamma_k \mat{A}_k^\dagger }{\gamma_k - i\omega}\right)^{-1}
\tfrac{\mat{B}_\ell^\dagger\gamma_\ell}{\gamma_\ell-i\omega}+ \tfrac{\mat{R}(\omega)}{\omega^2a_0^2}
\end{gather}
In order to evaluate the integral in \eqref{multi-phi-correlation}, we need to find the residues of 
\begin{gather}
\mat{f}(z)\equiv 
\mat{R}_{\tilde{\vecg{\phi}}\tilde{\vecg{\phi}}}(z)
(1-e^{iz t})
\end{gather}
Following similar steps as in Sec.~A.2.1, the residue at $\omega = 0$ gives
\begin{equation}
\mat{J}_0 =
\left[ \mat{B}\mat{A}^{-1}\tfrac{\mat{R}_0}{a_0^2}(\mat{B}\mat{A}^{-1})^\dagger + \tfrac{\mat{R}_0}{a_0^2}\right]t
\end{equation}
{ where we introduced the notation  $\mat{R}_0$ to denote the  diagonal autocorrelation matrix (Table 3 in the main text) evaluated at the lasing  frequency $\omega_\sigma$, i.e.,   $\mat{R}_0 \equiv \mat{R}(\omega_\sigma)$. } Near RO frequencies, we use the approximation for the autocorrelation [\eqref{multi-R-phiphi-approx}]
\begin{gather}
\mat{R}_{\tilde{\vecg{\phi}}\tilde{\vecg{\phi}}}(\omega) \approx
 \frac{1}{a_0^2}\cdot
\displaystyle\sum_{\substack{k\ell \\ \mu\nu\sigma\tau}}
\frac{\gamma_k\mat{B}_k}{\gamma_k + i\omega}
\frac{\mat{P}_{-\mu}\mat{P}_{+\nu}\mat{R}(\omega)\mat{P}^\dagger_{+\sigma}\mat{P}_{-\tau}}
{(\omega-\omega_{-\mu})(\omega - \omega_{+\nu})(\omega - \omega_{+\sigma}^*)(\omega-\omega_{-\tau}^*)}
 \frac{\gamma_\ell\mat{B}_\ell^\dagger}{\gamma_\ell - i\omega} + 
 \frac{\mat{R}(\omega)}{\omega^2 a_0^2}
  \end{gather}
  So the residues are 
 \begin{gather}
\mat{J}_\pm = 2 i\,\mathrm{Re}
\sum_\sigma 
\left[\mathrm{Res}(\mat{f},\omega_{+\sigma}) + \mathrm{Res}(\mat{f},\omega_{-\sigma})\right] =
\\
\,\mathrm{Re}\left\{\displaystyle\sum_{\substack{k\ell \\ \mu\nu\sigma\tau}}
{
\left(
\tfrac{
(1-e^{i\omega_{-\sigma} t})
(2i\gamma_k\gamma_\ell\mat{B}_k
\mat{P}_{-\sigma}\mat{P}_{+\nu} \mat{R}_-\mat{P}_{+\mu}^\dagger\mat{P}_{-\tau}^\dagger
\mat{B}^\dagger_\ell)}
{a_0^2\omega_{-\sigma}^2
(\omega_{-\sigma} - \omega_{+\nu})(\omega_{-\sigma} - \omega_{+\mu}^*)(\omega_{-\sigma} - \omega_{-\tau}^*)
}+
\tfrac{
(1-e^{i\omega_{+\sigma} t})
(2i\gamma_k\gamma_\ell\mat{B}_k
\mat{P}_{-\mu}\mat{P}_{+\sigma} \mat{R}_+\mat{P}_{+\nu}^\dagger\mat{P}_{-\tau}^\dagger
\mat{B}^\dagger_\ell)}
{a_0^2\omega_{+\sigma}^2
(\omega_{+\sigma} - \omega_{-\mu})(\omega_{+\sigma} - \omega_{+\nu}^*)(\omega_{+\sigma} - \omega_{-\tau}^*)
}
\right)
}
\right\}.
\end{gather}
For convenience, we rewrite the last results as 
\begin{align}
\mat{J}_\pm = 
\sum_{\sigma}\left[
\mat{S}^\sigma(1
-e^{-\Gamma_{\sigma}t}\cos\Omega_{\sigma}t)
+\mat{T}^\sigma e^{-\Gamma_{\sigma}t}\sin\Omega_{\sigma}t \right],
\end{align}
where we introduced
\begin{align}
&\mat{S}^\sigma\equiv
\sum_{k\ell\mu\nu\tau}
\mathrm{Re}
\left(
\tfrac{
2i\gamma_k\gamma_\ell\mat{B}_k
\mat{P}_{-\sigma}\mat{P}_{+\nu} \mat{R}_-\mat{P}_{+\mu}^\dagger\mat{P}_{-\tau}^\dagger
\mat{B}^\dagger_\ell}
{a_0^2\omega_{-\sigma}^2
(\omega_{-\sigma} - \omega_{+\nu})(\omega_{-\sigma} - \omega_{+\mu}^*)(\omega_{-\sigma} - \omega_{-\tau}^*)
}+
\tfrac{
2i\gamma_k\gamma_\ell\mat{B}_k
\mat{P}_{-\mu}\mat{P}_{+\sigma} \mat{R}_+\mat{P}_{+\nu}^\dagger\mat{P}_{-\tau}^\dagger
\mat{B}^\dagger_\ell}
{a_0^2\omega_{+\sigma}^2
(\omega_{+\sigma} - \omega_{-\mu})(\omega_{+\sigma} - \omega_{+\nu}^*)(\omega_{+\sigma} - \omega_{-\tau}^*)
}
\right),
 \nonumber\\
&\mat{T}^{\sigma}\equiv
\sum_{k\ell\mu\nu\tau}
\mathrm{Im}
\left(
\tfrac{
2i\gamma_k\gamma_\ell\mat{B}_k
\mat{P}_{-\mu}\mat{P}_{+\sigma} \mat{R}_+\mat{P}_{+\nu}^\dagger\mat{P}_{-\tau}^\dagger
\mat{B}^\dagger_\ell}
{a_0^2\omega_{+\sigma}^2
(\omega_{+\sigma} - \omega_{-\mu})(\omega_{+\sigma} - \omega_{+\nu}^*)(\omega_{+\sigma} - \omega_{-\tau}^*)
}
-\tfrac{
2i\gamma_k\gamma_\ell\mat{B}_k
\mat{P}_{-\sigma}\mat{P}_{+\nu} \mat{R}_-\mat{P}_{+\mu}^\dagger\mat{P}_{-\tau}^\dagger
\mat{B}^\dagger_\ell}
{a_0^2\omega_{-\sigma}^2
(\omega_{-\sigma} - \omega_{+\nu})(\omega_{-\sigma} - \omega_{+\mu}^*)(\omega_{-\sigma} - \omega_{-\tau}^*)
}\right)
\end{align} 
and 
{  we introduced the notation  $\mat{R}_\pm$ to denote the   autocorrelation matrix  evaluated at the RO   frequency $\omega_\sigma\pm\Omega$. Note that $\mat{R}_+ \approx \mat{R}_-$ since the gain is symmetric around the lasing frequencies.} 
Collecting the terms, we find that the phase variance is 
\begin{subequations}
\begin{empheq}[box=\widefbox]{align}
&\left<\Scale[0.95]{
[\phi_\mu(t) - \phi_\mu(0)] [\phi_\nu(t) - \phi_\nu(0)]}\right> =\nonumber \\
&\left(
\frac{[\mat{B}\mat{A}^{-1}\mat{R}_0{\mat{A}^\dagger}^{-1}\mat{B}^\dagger]_{\mu\nu}}{a_0^2}
+
\frac{[\mat{R}_0]_{\mu\nu}}{a_0^2}
\right)2t+
\sum_{\sigma}\left[
\mat{S}_{\mu\nu}^\sigma(1
-e^{-\Gamma_{\sigma}t}\cos\Omega_{\sigma}t)
+\mat{T}_{\mu\nu}^\sigma e^{-\Gamma_{\sigma}t}\sin\Omega_{\sigma}t \right]\nonumber
\end{empheq}
\end{subequations}

\section*{B.3.2 Intensity autocorrelation}

In a similar manner, we can also obtain the multimode intensity autocorrelations. 
As in the single-mode case, we need to compute
\begin{gather}
\left<[\vec{u}(t+t') + \vec{u}(t')][\vec{u}^T\!(t+t') + \vec{u}^T\!(t')]\right>=
\pi^{-1}\mathrm{Re}\int_{-\infty}^\infty d\omega\,\mat{R}_{\tilde{\vec{u}}\tilde{\vec{u}}}(\omega) (1+e^{i\omega t}) \equiv\mat{G}_\pm
\end{gather}
Denoting  the integrand   by
\begin{equation}
\mat{f}(z) = \mat{R}_{\tilde{\vec{u}}\tilde{\vec{u}}}(z)(1+e^{iz t}),
\end{equation}
the autocorrelation is
\begin{gather}
\mat{G}_\pm=
 2 i\mbox{Re} \left[\sum_\sigma 
\mathrm{Res}(\mat{f},\omega_{+\sigma}) + \mathrm{Res}(\mat{f},\omega_{-\sigma})\right]
\end{gather}
The integrand only has poles near the RO frequencies. We use the approximation [\eqref{multi-R-uu-approx}]:
\begin{equation}
\mat{R}_{\tilde{\vec{u}}\tilde{\vec{u}}}(\omega) \approx 
\frac{\omega^2}{a_0^2}\cdot
\displaystyle\sum_{ \mu\nu\sigma\tau}
\frac{\mat{P}_{-\mu}\mat{P}_{+\nu}\mat{R}(\omega)\mat{P}^\dagger_{+\sigma}\mat{P}_{-\tau}}
{(\omega-\omega_{-\mu})(\omega - \omega_{+\nu})(\omega - \omega_{+\sigma}^*)(\omega-\omega_{-\tau}^*)}
\end{equation}
Next, we perform the integration using Cauchy's theorem and obtain
\begin{align}
\mat{G}_\pm &= 
 \mathrm{Re}\left\{\sum_{\mu\nu\sigma\tau}
 \tfrac{2i}
{a_0^2}
\left(\tfrac{\mat{P}_{-\sigma}\mat{P}_{+{\mu}} \mat{R}(\omega_-)\mat{P}_{+\nu}^\dagger\mat{P}_{-\tau}^\dagger(1+e^{i\omega_{-\sigma} t})\omega_{-\sigma}^2}{(\omega_{-\sigma} -\omega_{+{\mu}})(\omega_{-\sigma} - \omega_{-\nu}^*)(\omega_{-\sigma} - \omega_{+{\tau}}^*)}+
\tfrac{\mat{P}_{-\mu}\mat{P}_{+{\sigma}} \mat{R}(\omega_+)\mat{P}_{+\nu}^\dagger\mat{P}_{-\tau}^\dagger(1+e^{i\omega_{+{\sigma}} t})\omega_{+\sigma}^2}{(\omega_{+{\sigma}} - \omega_{-\mu}) (\omega_{+{\sigma}} - \omega_{-\nu}^*)(\omega_{+{\sigma}} - \omega_{+{\tau}}^*)}\right)\right\}
\end{align}
Once again, we rewrite the result in compact form as
\begin{align}
\boxed{
\Scale[0.95]{
\left<[u_\mu(t) + u_\mu(0)][u_\nu(t) + u_\nu(0)]\right>=
\sum_{\sigma}\left[
\mat{U}^\sigma_{\mu\nu}(1
+e^{-\Gamma_{\sigma}t}\cos\Omega_{\sigma}t)
+\mat{V}^\sigma_{\mu\nu} e^{-\Gamma_{\sigma}t}\sin\Omega_{\sigma}t \right],}}
\end{align}
where we introduced the matrices
\begin{subequations}
\begin{gather}
\mat{U}^\sigma\equiv\mathrm{Re}\left\{\sum_{\mu\nu\tau}\tfrac{
2i\omega_{-\sigma}^2\mat{P}_{-\sigma}\mat{P}_{+{\mu}} \mat{R}_-\mat{P}_{+\nu}^\dagger\mat{P}_{-\tau}^\dagger}
{a_0^2(\omega_{-\sigma} -\omega_{+{\mu}})(\omega_{-\sigma} - \omega_{-\nu}^*)(\omega_{-\sigma} - \omega_{+{\tau}}^*)}
+
 \tfrac{2i\omega_{+\sigma}^2\mat{P}_{-\mu}\mat{P}_{+{\sigma}} \mat{R}_+\mat{P}_{+\nu}^\dagger\mat{P}_{-\tau}^\dagger}
 {a_0^2(\omega_{+{\sigma}} - \omega_{-\mu}) (\omega_{+{\sigma}} - \omega_{-\nu}^*)(\omega_{+{\sigma}} - \omega_{+{\tau}}^*)}\right\}
\\
\mat{V}^\sigma\equiv-\mathrm{Im}\left\{\sum_{\mu\nu\tau}\tfrac{
2i\omega_{-\sigma}^2\mat{P}_{-\sigma}\mat{P}_{+{\mu}} \mat{R}_-\mat{P}_{+\nu}^\dagger\mat{P}_{-\tau}^\dagger}
{a_0^2(\omega_{-\sigma} -\omega_{+{\mu}})(\omega_{-\sigma} - \omega_{-\nu}^*)(\omega_{-\sigma} - \omega_{+{\tau}}^*)}
-
 \tfrac{2i\omega_{-\sigma}^2\mat{P}_{-\mu}\mat{P}_{+{\sigma}} \mat{R}_+\mat{P}_{+\nu}^\dagger\mat{P}_{-\tau}^\dagger}
 {a_0^2(\omega_{+{\sigma}} - \omega_{-\mu}) (\omega_{+{\sigma}} - \omega_{-\nu}^*)(\omega_{+{\sigma}} - \omega_{+{\tau}}^*)}\right\}.
\end{gather}
\end{subequations}

\section*{B.3.3 The cross term}


Finally, we compute the multimode cross term
\begin{align}
\langle[\vecg{\phi}(t+t') - \vecg{\phi}(t')][\vec{u}^T(t+t') + \vec{u}^T(t')]\rangle=
\frac{1}{2\pi}\int_{-\infty}^\infty \!\!{d}\omega \, \left(e^{i\omega t} - e^{-i\omega t}\right)
\,\mat{R}_{\tilde{\vecg{\phi}}\tilde{\vec{u}}}(\omega)
\equiv \mat{I}_0+\mat{I}_\pm.
\end{align}
The multimode phase-intensity autocorrelation is given by \eqref{multi-R-phiu},
\begin{equation}
\Scale[0.9]{\mat{R}_{\tilde{\vecg{\phi}}\tilde{\vec{u}}}(\omega) = 
\sum_{k} \tfrac{\gamma_k\mat{B}_k}{\gamma_k + i\omega}\cdot
\left({i\omega\mathbb{1} + \sum_k  \tfrac{\gamma_k \mat{A}_k }{\gamma_k+ i\omega}
}\right)^{-1}\frac{\mat{R}(\omega)}{i\omega a_0^2}
\left({-i\omega\mathbb{1} + \sum_k  \tfrac{\gamma_k \mat{A}_k^\dagger }{\gamma_k- i\omega}
}\right)^{-1}}.
\end{equation}
We define the integrand as 
\begin{equation}
\mat{f}(z)\equiv
2i\sin(zt)
\,\mat{R}_{\tilde{\vecg{\phi}}\tilde{\vec{u}}}(z).
\end{equation}
The residue at zero gives
\begin{equation}
\mat{I}_0 = 2\pi i\, \mathrm{Res}(\mat{f},0) = 
\frac{1}{a_0^2}\mat{B}\mat{A}^{-1}\mat{R}_0(\mat{A}^{-1})^{\dagger}.
\end{equation}
For the RO-related terms, we use the approximation [\eqref{multi-R-phiu-approx}]:
\begin{equation}
\mat{R}_{\tilde{\vecg{\phi}}\tilde{\vec{u}}}(\omega) \approx 
\sum_{k\mu\nu\sigma\tau}
\left[\frac{\gamma_k }{i\omega}-1\right]
\frac{\gamma_k\mat{B}_k}{\gamma_k^2 + \omega^2}
\frac{\mat{P}_{-\mu}\mat{P}_{+\nu}\mat{R}(\omega)\mat{P}^\dagger_{+\sigma}\mat{P}_{-\tau}}
{(\omega-\omega_{-\mu})(\omega - \omega_{+\nu})(\omega - \omega_{+\sigma}^*)(\omega-\omega_{-\tau}^*)}\cdot 
\frac{1}{ a_0^2} 
\label{eq:drop-1}
\end{equation}
Now we compute the residues in order to find
\begin{equation}
\mat{I}_\pm = 2\pi i\,\mathrm{Re}
\sum_\sigma 
\left[\mathrm{Res}(\mat{f},\omega_{+\sigma}) + \mathrm{Res}(\mat{f},\omega_{-\sigma})\right]
\end{equation}
When computing the residues at $\omega_\pm$, 
we drop the $1$ inside the square brackets in \eqref{drop-1} [changing $\left(\frac{\gamma_k }{i\omega}-1\right)$ to $\frac{\gamma_k }{i\omega}$], because the integrand is $\sin(zt)\mat{R}_{\tilde{\vecg{\phi}}\tilde{\vec{u}}}(z) $ and $\sin$ is odd so   only the odd part of $\mat{R}_{\tilde{\vecg{\phi}}\tilde{\vec{u}}}$ gives a non-zero contribution. Moreover, we approximate $\gamma_k+\omega^2\approx \omega^2$, which holds near the RO frequencies.
 We find
\begin{align}
&\mat{I}_\pm =
\Scale[1.1]{ \displaystyle\sum_{k\mu\nu\sigma\tau}\tfrac{1}{a_0^2}
\left(\tfrac{
2\gamma_k^2\mat{B}_k\,\mat{P}_{-\sigma}\mat{P}_{+{\mu}} \mat{R}_-\mat{P}_{+\nu}^\dagger\mat{P}_{-{\tau}}^\dagger
e^{i\omega_{-\sigma} t}}
{\omega_{-\sigma}(\omega_{-\sigma} - \omega_{+{\mu}})(\omega_{-\sigma} - \omega_{+\nu}^*)(\omega_{-\sigma} - \omega_{-{\tau}}^*)}+
\tfrac{2\gamma_k^2\mat{B}_k\,\mat{P}_{-\mu}\mat{P}_{+{\sigma}} \mat{R}_+\mat{P}_{+\nu}^\dagger\mat{P}_{-{\tau}}^\dagger
e^{i\omega_{+\sigma} t}}{\omega_{+\sigma}(\omega_{+\sigma} - \omega_{-\mu}) (\omega_{+\sigma} - \omega_{+\nu}^*)(\omega_{+\sigma} -\omega_{-\tau}^*)}\right),}
\end{align}
which can be rewritten as 
\begin{subequations}
\begin{empheq}[box=\widefbox]{align}
&\left<[u_\mu(t) + u_\mu(0)][\phi_\nu(t) + \phi_\nu(0)]\right>=\nonumber \\
&\tfrac{\left[2\mat{B}\mat{A}^{-1}\mat{R}_0\mat{A}^{-1}\right]_{\mu\nu}}{a_0^2}+
\sum_{\sigma}\left[
\mat{X}^\sigma_{\mu\nu} e^{-\Gamma_{\sigma}t}\cos\Omega_{\sigma}t
+\mat{Y}^\sigma_{\mu\nu} e^{-\Gamma_{\sigma}t}\sin\Omega_{\sigma}t \right],
\end{empheq}
\end{subequations}
where we introduced the definitions
\begin{align}
&\mat{X}^\sigma\equiv
\Scale[1.1]{ \displaystyle\sum_{k\mu\nu\tau}\tfrac{1}{a_0^2}
\left(\tfrac{
2\gamma_k^2\mat{B}_k\,\mat{P}_{-\sigma}\mat{P}_{+{\mu}} \mat{R}_-\mat{P}_{+\nu}^\dagger\mat{P}_{-{\tau}}^\dagger
}
{\omega_{-\sigma}(\omega_{-\sigma} - \omega_{+{\mu}})(\omega_{-\sigma} - \omega_{+\nu}^*)(\omega_{-\sigma} - \omega_{-{\tau}}^*)}+
\tfrac{2\gamma_k^2\mat{B}_k\,\mat{P}_{-\mu}\mat{P}_{+{\sigma}} \mat{R}_+\mat{P}_{+\nu}^\dagger\mat{P}_{-{\tau}}^\dagger}{\omega_{+\sigma}(\omega_{+\sigma} - \omega_{-\mu}) (\omega_{+\sigma} - \omega_{+\nu}^*)(\omega_{+\sigma} -\omega_{-\tau}^*)}\right),}
\nonumber\\
&\mat{Y}^\sigma\equiv
\Scale[1.1]{ \displaystyle\sum_{k\mu\nu\tau}\tfrac{i}{a_0^2}
\left(\tfrac{
2\gamma_k^2\mat{B}_k\,\mat{P}_{-\sigma}\mat{P}_{+{\mu}} \mat{R}_-\mat{P}_{+\nu}^\dagger\mat{P}_{-{\tau}}^\dagger
}
{\omega_{-\sigma}(\omega_{-\sigma} - \omega_{+{\mu}})(\omega_{-\sigma} - \omega_{+\nu}^*)(\omega_{-\sigma} - \omega_{-{\tau}}^*)}
-
\tfrac{2\gamma_k^2\mat{B}_k\,\mat{P}_{-\mu}\mat{P}_{+{\sigma}} \mat{R}_+\mat{P}_{+\nu}^\dagger\mat{P}_{-{\tau}}^\dagger}{\omega_{+\sigma}(\omega_{+\sigma} - \omega_{-\mu}) (\omega_{+\sigma} - \omega_{+\nu}^*)(\omega_{+\sigma} -\omega_{-\tau}^*)}\right)}.
\end{align}

\section*{B.4. From second-order moments to the multimode autocorrelations }

In the previous section, we found that the second-order moments have the form
\begin{subequations}
 \begin{gather}
\Scale[0.9]{\left<[\vecg{\phi}(t)-\vecg{\phi}(0)][\vecg{\phi}^T(t)-\vecg{\phi}^T(0)]\right>  =
 \mat{Q}^{(1)} t+
  \sum_\sigma \left[ \mat{Q}^{(2)}_{\sigma}(1-e^{-\Gamma_\sigma t}\cos\Omega_\sigma t) + \mat{Q}^{(3)}_{\sigma}e^{-\Gamma_\sigma t}\sin\Omega_\sigma t\right]}
\label{eq:1}
\\ 
\left<[\vec{u}(t)+\vec{u}(0)][\vec{u}^T(t)+\vec{u}^T(0)]\right>  = 
\sum_\sigma \mat{Q}^{(4)}_{\sigma}(1+e^{-\Gamma_\sigma t}\cos\Omega_\sigma t)  + \mat{Q}^{(5)}_{\sigma} e^{-\Gamma_\sigma t}\sin\Omega_\sigma t
\label{eq:2}
\\ 
\left<[\vec{u}(t)+\vec{u}(0)][\vecg{\phi}^T(t)-\vecg{\phi}^T(0)]\right>  = 
\mat{Q}^{(6)}+ \sum_\sigma \mat{Q}^{(7)}_{\sigma}e^{-\Gamma_\sigma t}\cos\Omega_\sigma t + \mat{Q}^{(8)}_{\sigma}e^{-\Gamma_\sigma t}\sin\Omega_\sigma t
\label{eq:3}
\end{gather}
\label{eq:multi-correlations-formal}
\end{subequations}
\hspace{-3pt}Comparing the boxed equations with multi-correlations-formal, we find:
\begin{gather}
\mat{Q}^{(1)} = 2\left(\frac{\mat{B}\mat{A}^{-1}\mat{R}_0{\mat{A}^\dagger}^{-1}\mat{B}^\dagger}{a_0^2}+\frac{\mat{R}_0}{a_0^2}\right),\quad
\mat{Q}^{(2)}_{\sigma} = \mat{S}^\sigma,\quad
\mat{Q}^{(3)}_{\sigma} = \mat{T}^\sigma,\quad
\mat{Q}^{(4)}_{\sigma} = \mat{U}^\sigma,\quad
\mat{Q}^{(5)}_{\sigma} = \mat{V}^\sigma,\nonumber\\
\mat{Q}^{(6)}= \tfrac{2\mat{B}\mat{A}^{-1}\mat{R}_)\mat{A}^{-1}}{a_0^2},\quad
\mat{Q}^{(7)}_{\sigma}= \mat{X}^\sigma,\quad
\mat{Q}^{(8)}_{\sigma} = \mat{Y}^\sigma.
\end{gather}
Following similar steps as in the single-mode regime, one can show that in the limit of strong phase diffusion [see discussion following \eqref{TransformMe} for quantitative definition], the Fourier transform of the multimode autocorrelation takes the form
\begin{gather}
\frac{1}{\sqrt{2\pi}}\int_{-\infty}^\infty 
dt e^{-i\omega t} \langle a_\mu(t) a_\nu^*(0) \rangle= 
\underbrace{\tfrac{\mat{Q}^{(1)}_{\mu\nu}}{(\omega - \omega_\mu)^2+(\mat{Q}^{(1)}_{\mu\nu}/2)^2}
\left(1-\tfrac{\sum_\sigma \left\{\mat{Q}^{(2)}_{\mu\nu\sigma}+\mat{Q}^{(4)}_{\mu\nu\sigma}\right\}}{2}\right)}_{\text{central peaks}} +\nonumber\\
\underbrace{
\sum_\sigma
\tfrac{\Gamma_\mathrm{SB}^{\mu\nu\sigma}}{((\omega - \omega_\mu)+\Omega_\sigma)^2+(\Gamma_\mathrm{SB}^{\mu\nu\sigma})^2}
\left[
\left(\tfrac{ \mat{Q}^{(2)}_{\mu\nu\sigma}+\mat{Q}^{(4)}_{\mu\nu\sigma}+2\mat{Q}^{(8)}_{\mu\nu\sigma}}{2}\right)
+
\tfrac{\Omega_\sigma+(\omega - \omega_\mu)}{\Gamma_\mathrm{SB}^{\mu\nu\sigma}}
\left(\tfrac{\mat{Q}^{(5)}_{\mu\nu\sigma}-\mat{Q}^{(3)}_{\mu\nu\sigma}+2\mat{Q}^{(7)}_{\mu\nu\sigma}}{2}\right)
\right]}_{\text{blue sidepeaks}}+\nonumber\\
\underbrace{\sum_\sigma\tfrac{\Gamma_\mathrm{SB}^{\mu\nu\sigma}}{((\omega - \omega_\mu)-\Omega_\sigma)^2+(\Gamma_\mathrm{SB}^{\mu\nu\sigma})^2}
\left[
\left(\tfrac{\mat{Q}^{(2)}_{\mu\nu\sigma}+\mat{Q}^{(4)}_{\mu\nu\sigma}-2\mat{Q}^{(8)}_{\mu\nu\sigma}}{2}\right)
-
\tfrac{\Omega_\sigma-(\omega - \omega_\mu)}{\Gamma_\mathrm{SB}^{\mu\nu\sigma}}
\left(\tfrac{\mat{Q}^{(5)}_{\mu\nu\sigma}-\mat{Q}^{(3)}_{\mu\nu\sigma}-2\mat{Q}^{(7)}_{\mu\nu\sigma}}{2}\right)
\right]}_{\text{red  sidepeaks}}.
\end{gather} 
where $\Gamma_\mathrm{SB}^{\mu\nu\sigma} = \frac{\Gamma_{\mu\nu}}{2}+\Gamma_\sigma$. 
This completes the derivation of the multimode noise spectrum.






\end{document}